\documentclass[fleqn,usenatbib]{mnras}

\usepackage{newtxtext,newtxmath}

\usepackage[T1]{fontenc}
\usepackage{ae,aecompl}

\DeclareRobustCommand{\VAN}[3]{#2}
\let\VANthebibliography\thebibliography
\def\thebibliography{\DeclareRobustCommand{\VAN}[3]{##3}\VANthebibliography}

\usepackage{graphicx}	
\usepackage{amsmath}	

\DeclareMathSymbol{\mlq}{\mathord}{operators}{``}
\DeclareMathSymbol{\mrq}{\mathord}{operators}{`'}

\title[Transmission strings]{Transmission strings: a technique for spatially mapping exoplanet atmospheres around their terminators}

\author[D. Grant and H.R. Wakeford]{
David Grant$^{1}$\thanks{E-mail: david.grant@bristol.ac.uk}
and Hannah R. Wakeford$^{1}$
\\
$^{1}$University of Bristol, HH Wills Physics Laboratory, Tyndall Avenue, Bristol, BS8 1TL, UK\\
}

\date{Accepted 3 December 2022. Received 1 December 2022; in original form 27 October 2022}

\pubyear{2022}

\begin{document}
\label{firstpage}
\pagerange{\pageref{firstpage}--\pageref{lastpage}}
\maketitle

\begin{abstract}
Exoplanet transmission spectra, which measure the absorption of light passing through a planet’s atmosphere during transit, are most often assessed globally, resulting in a single spectrum per planetary atmosphere. However, the inherent three-dimensional nature of planetary atmospheres, via thermal, chemical, and dynamical processes, can imprint inhomogeneous structure and properties in the observables. In this work, we devise a technique for spatially mapping the atmospheres of exoplanets in transmission. Our approach relaxes the assumption that transit light curves are created from circular stars occulted by circular planets, and instead we allow for flexibility in the planet’s sky-projected shape. We define the planet's radius to be a single-valued function of angle around its limb, and we refer to this mathematical object as a {\it transmission string}. These transmission strings are parameterised in terms of Fourier series, a choice motivated by these series having adjustable complexity, generating physically practical shapes, while being reducible to the classical circular case. The utility of our technique is primarily intended for high-precision multi-wavelength light curves, from which inferences of transmission spectra can be made as a function of angle around a planet's terminator, enabling analysis of the multidimensional physics at play in exoplanet atmospheres. More generally, the technique can be applied to any transit light curve to derive the shape of the transiting body. The algorithm we develop is available as an open-source package, called \texttt{harmonica} \footnotemark{}.
\end{abstract}

\begin{keywords}
exoplanets -- planets and satellites: atmospheres -- software: data analysis -- techniques: spectroscopic -- techniques: photometric
\end{keywords}



\section{Introduction}
\label{sec:tracing_introduction}

\renewcommand*{\thefootnote}{\fnsymbol{footnote}}
\footnotetext[2]{\url{https://github.com/DavoGrant/harmonica}}
\footnotetext[2]{\url{https://harmonica.readthedocs.io}}

Planets are inherently three-dimensional (3D) objects, with radiative and advective timescales governed by 3D chemical and dynamical processes. To study these processes, general circulation models (GCMs) have been developed to simulate the multidimensional physics at play in the atmospheres of exoplanets \citep[e.g.,][]{Showman2009, menou2009atmospheric, mayne2014unified}. These GCMs can help inform us of how physical processes may be manifested in observations, and thereby how the physics may become testable. 

For hot Jupiters, tidally-locked orbits and close proximity to their host stars create permanent, strong day-night heating contrasts. This process is a key driver of the resulting atmospheric circulation \citep{showman2013atmospheric, pierrehumbert2019atmospheric, zhang2020atmospheric}, leading to rotational and divergent flows \citep{hammond2021rotational} and net easterly equatorial superrotation \citep{showman2010matsuno}. For these types of planets, several observable multidimensional effects have been theorised. These include non-uniform thermal structure \citep{feng2016impact, KomacekShowman2016, KomacekShowmanTan2017, taylor2020understanding}, temperature and molecular abundance differences between leading and trailing limbs \citep{fortney2010}, transit timing offsets relating to differing limb signatures \citep{dobbs-dixon2012}, advection of the hot spot downwind of the substellar point \citep{cooper2005dynamic, Burrows2010ApJ, kataria2015atmospheric, parmentier2018exoplanet}, and inhomogeneous cloud and haze coverage at the terminator \citep{kempton2017, Powell2019ApJ, Lines2018HD209, Lines2019HD189}. For smaller terrestrial planets, although the observables are significantly weaker, processes such as the presence of a surface \citep{may2020super} or the exact physical prescription \citep{sergeev2022bistability} can influence the atmospheric circulation.

\begin{figure*}
\centering
\includegraphics[width=\textwidth]{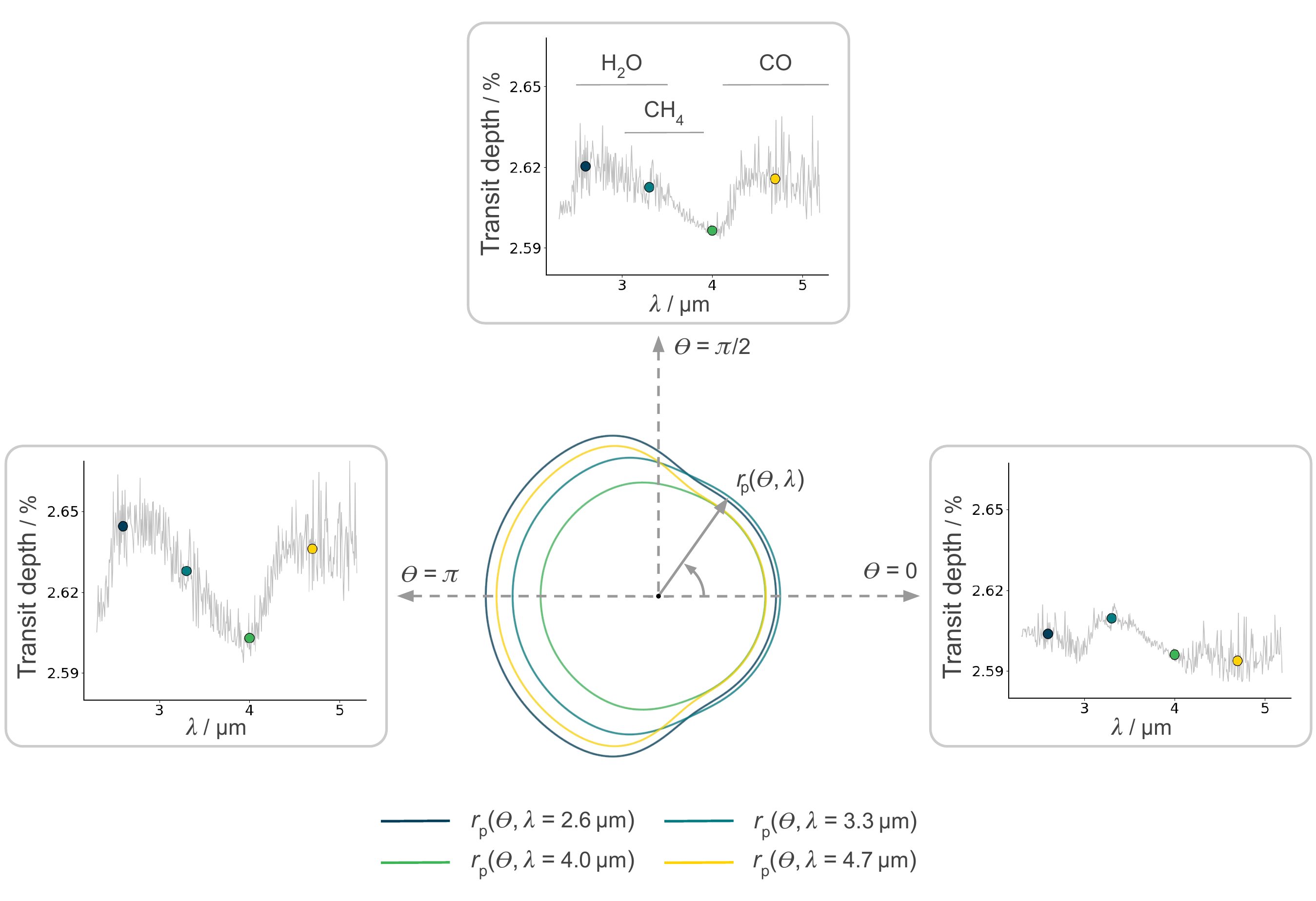}
\caption{Concept schematic of transmission strings. Centre: four example transmission strings inferred from four light curve fits at wavelengths of 2.6, 3.3, 4.0, and 4.7 \textmu m. The eastern (trailing) terminator is on the left-hand side. Note that the deviations from circular have been amplified for visual clarity. Surrounding panels: three cuts around the terminator at angles of 0, $\pi/2$, and $\pi$ radians, produce three different transmission spectra. The transmission spectra are shown in units of transit depth, defined as $r_{\rm{p}}^2 / R_{\rm{s}}^2$ where $R_{\rm{s}}$ is the radius of the star. The transit depth data points correspond in colour with the transmission strings. These data are based on the models of \citet{fortney2010}, and show the transition in $\rm{CO/CH}_4$ abundance ratio from the cooler western terminator to the hotter eastern terminator.}
\label{fig:transmission_string_schematic}
\end{figure*}

\begin{figure}
\centering
\includegraphics[width=\columnwidth]{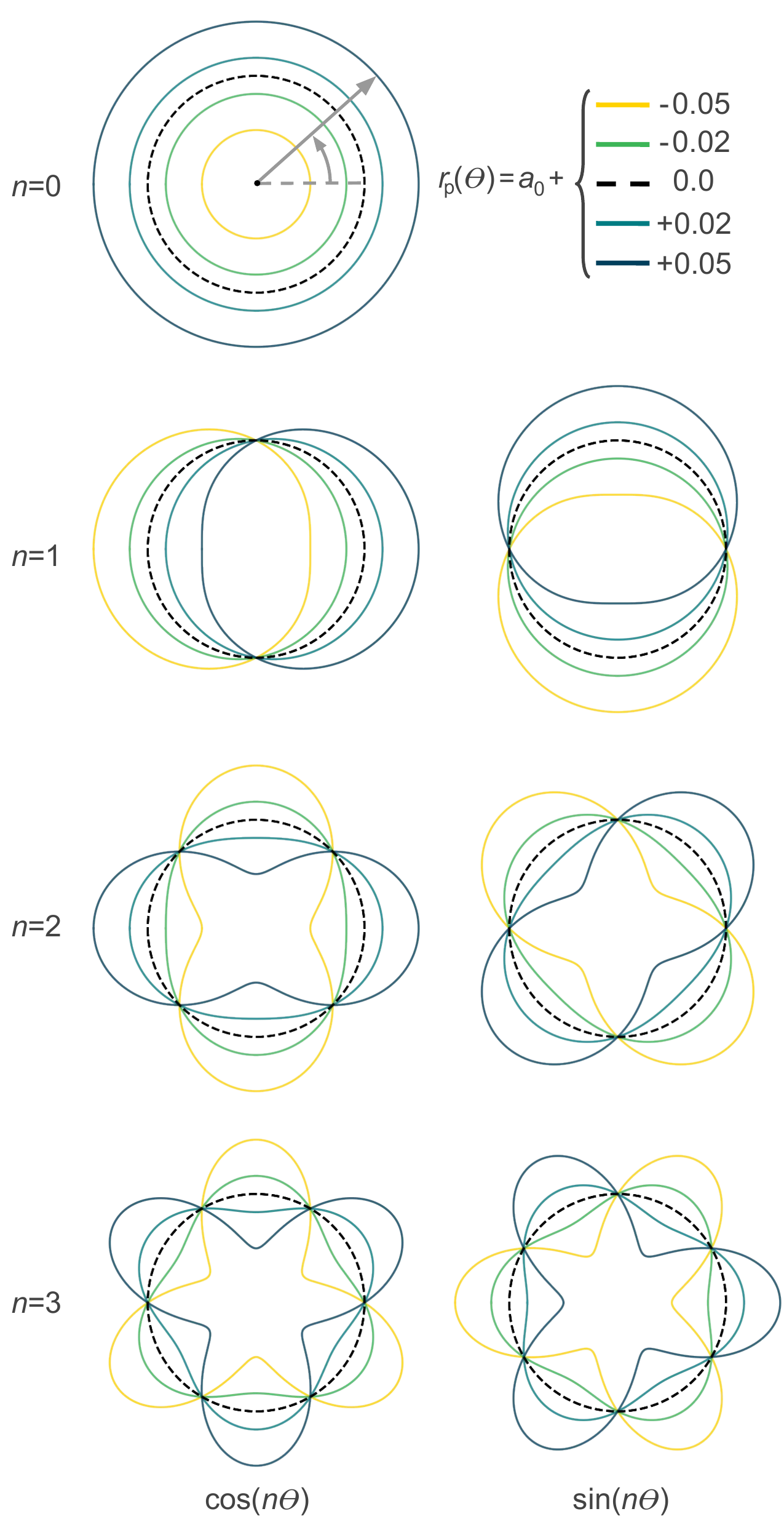}
\caption{The first 7 terms of our Fourier parameterisation of transmission strings. Each shape is the sum of the zeroth order term, $a_0=0.1$, and the corresponding grid point's high order term at varying amplitudes. For example, the bottom-left shape is given by $r_{\rm{p}}(\theta) = 0.1 + a_3 \cos{3 \theta}$, where $a_3 \in \{ -0.05, -0.02, 0.0, 0.02, 0.05 \}$.}
\label{fig:fourier_terms_schematic}
\end{figure}

From the observational side, evidence for some of these multidimensional effects has already been found. Longitudinal variations, and hot-spot offsets, have been observed from numerous phase curves \citep[e.g.,][]{cowan2012thermal, Demory2013ApJ, Beatty2019AJ, Keating2019NatAs, mikal2022diurnal}, dayside brightness distributions measured via eclipse mapping \citep{knutson2007map, de2012towards, majeau2012two}, and high-resolution ground-based observations have revealed insights into the condensation of species across the day-night transition \citep[e.g.,][]{Ehrenreich2020, kesseli2022atomic}. However, these measurements are hard to make and require precise high-cadence observations limiting investigations to the hottest and brightest targets.

With the launch of JWST, transit data is now available with greater precision than ever before. Previous analysis into multidimensional observables has centred around phase curves. But, for some planets the planet-to-star contrast ratio may be more favourable in transmission; and therefore, the multidimensional signals more susceptible to detection. In fact, it was shown by \citet{MacDonald2020} that accounting for the multidimensional nature of planetary atmospheres in transmission is vital to producing unbiased inferences. To extract these signals in transmission, one potential method is to fit light curves with a circular occultation model \citep[e.g.,][]{mandel2002analytic}, and then, with the resulting transmission spectrum perform multidimensional retrieval modelling \citep[e.g.,][]{MacDonaldLewis2022}. A transmission spectrum produced in this way constitutes some overall average of the atmospheric conditions through which the light propagates, and thus requires the retrievals to disentangle any multidimensional variability. Inferring differences around the terminator, along the line of sight, and in elevation, wrapped up in one spectrum, is a challenging modelling task.  

As shown by \citet{von2016inferring} and \citet{EspinozaJones2021AJ}, an alternative method is to draw out any spatial information available at the light curve fitting stage, prior to retrieval modelling. At both ingress and egress a planet only partially occults the star, with different sectors around the terminator interposing our line of sight at different times. Similarly, during transit the star's limb-darkened surface provides a variable backlight to different sectors of the atmosphere. These two effects mean that spatial information around the terminator is present directly in the light curve data. Therefore, it may be advantageous to draw out any variation around the terminator at the light-curve fitting stage, and alleviate multidimensional retrievals of one of their degrees of freedom.

To this end, we devise a general mathematical object for inferring atmospheric variability around the terminator directly from the light curve data. We refer to these objects as {\it transmission strings}. The utility of transmission strings is to enable the extraction of transmission spectra as a function of angle around a planet's terminator. This method utilises the spatial information as early as possible in the data analysis procedure, with the intention of making multidimensional retrieval modelling easier and improving our 3D inferences of exoplanet atmospheres.

Our study is structured as follows. In Section \ref{sec:transmission_strings} we describe the concept behind transmission strings and define their parameterisation in terms of Fourier series. In Section \ref{sec:computing_light_curves} we detail the mathematics and algorithm for computing the light curve associated with a given transmission string. In Section \ref{sec:performance_benchmarks} we measure the performance of our algorithm. In Section \ref{sec:demo} we demonstrate our technique's capabilities for inference on a JWST-like dataset. In Section \ref{sec:discussion} we discuss further complexities of our approach. Finally, in Section \ref{sec:summary_and_conclusions} we summarise our findings.

\section{Transmission strings}
\label{sec:transmission_strings}

Typically, a model fit to a transit light curve yields a single measurement of planet radius, $r_{\rm{p}}$, based on a model of a circular occultor. Here, we extend this modelling approach beyond circular shapes, allowing for variability around the planet's terminator. We define the planet radius to be a single-valued function of angle around the terminator, $r_{\rm{p}}(\theta)$, and we refer to this mathematical object as a transmission string. 

The utility of a transmission string is borne out by observations of light curves at multiple wavelengths. For each light curve a different transmission string may be inferred. From these transmission strings a transmission spectrum may be produced for any $\theta$. To illustrate this concept, in Figure \ref{fig:transmission_string_schematic} we display an example based on the models of \citet{fortney2010}. In these models of tidally-locked hot Jupiters, a common feature that appears is a wide low-latitude wind circulating from west to east. This circulation displaces the hottest and coldest regions of the planet, with the hot substellar atmosphere being advected towards the east. The result is a difference in temperature between the eastern and western terminators; and therefore, a possible difference in chemical abundances. If observed light curves from a planet of this nature are fit with a circular transit model, the resulting transmission spectrum averages out the variability, and may make the multidimensional information difficult to retrieve. 

Instead, if at the light curve fitting stage we measure a transmission string for each wavelength, as displayed in the centre of Figure \ref{fig:transmission_string_schematic}, we can generate different transmission spectra at each angle around the terminator, which show the planet's atmospheric variation. Specific to this example, the $\rm{CO/CH}_4$ abundance ratio shows a strong dependence on the temperature variation around the terminator. In the right-hand panel, the transmission spectrum at $\theta=0$ has prominent $\rm{CH}_4$ absorption at 3.3 \textmu m, but little $\rm{CO}$ absorption at 4.7 \textmu m. In contrast, the left-hand panel, showing the transmission spectrum at $\theta=\pi$, has weaker $\rm{CH}_4$ but far stronger $\rm{CO}$ absorption. Inspecting the difference between these two transmission strings, at 3.3 \textmu m (light blue) and at 4.7 \textmu m (yellow), we see how the 4.7 \textmu m transmission string infers a relatively smaller planet radius at $\theta=0$, but for angles around $\theta=\pi$ this radius has inflated beyond that of the 3.3 \textmu m transmission string. Between these wavelengths, at 4.0 \textmu m the transmission string (green) is fairly circular. This transmission string probes wavelengths between strong molecular absorption bands, deeper in the atmosphere, and is only slightly inflated around $\theta=\pi$ due to the temperature increasing the atmospheric scale height \citep{dobbs-dixon2012}.

The example above expresses how measurements of transmission strings, directly from the light curves, can extract multidimensional spectral information. Whilst this example focused on carbon chemistry, there are many further potential applications of transmission strings in investigations into the 3D chemical and dynamical processes in exoplanet atmospheres.

\subsection{Harmonic transmission strings}
\label{sec:harmonic_transmission_strings}

A transmission string may be parameterised by any single-valued function, $r_{\rm{p}}(\theta)$. Using this definition, the classical circular model of a transiting planet can be thought of as a transmission string where $r_{\rm{p}}(\theta)=\rm{constant}$. Another previous parameterisation includes the back-to-back semi-circles of \citet{von2016inferring} and \citet{EspinozaJones2021AJ}, equivalent to a transmission string parameterised by a top-hat function. However, this parameterisation assumes a rigid dichotomy between the east and west hemispheres of a planet, while 3D effects likely imprint continuous functions around the terminator of a planet's atmosphere \citep[e.g.,][]{dobbs-dixon2012, parmentier20133d}.

In this study, we expand the set of possible transmission strings into a more general framework. We define a transmission string in terms of a Fourier series, such that
\begin{equation}
r_{\rm{p}}(\theta) = \sum_{n=0}^{N_c} a_{n} \cos{(n \theta)} + \sum_{n=1}^{N_c} b_{n} \sin{(n \theta)},
\label{eq:transmission_string_real}
\end{equation}
where $a_n$ and $b_n$ are the $n$th harmonics' amplitudes, the total number of terms is equal to $2N_c + 1$, and $r_{\rm{p}}$ is in units of stellar radius. The angle $\theta$ is measured around the terminator, anti-clockwise in the plane of the sky, from the direction of the planet's sky-projected orbital velocity.

The first 7 terms of our parameterisation are displayed in Figure \ref{fig:fourier_terms_schematic}. The zeroth order term, $a_0$, is displayed at the top and represents the mean radius of the planet. Subsequent drawings show the deviations from a circular geometry generated by the high order terms. The value of this parameterisation lies in the flexibility to build transmission strings of arbitrary complexity, whilst being reducible to the classical circular case. For a model utilising only the first 4 terms, a transmission string may already encode variation of the mean radius ($a_0$), east-to-west differences ($a_1\cos{\theta}$), north-to-south differences ($b_1\sin{\theta}$), and equatorial-vs-polar inflation ($a_2\cos{2\theta}$). In reality, the complexity we are able to infer in a given transmission string will be driven by the quality of the data.

\section{Computing light curves}
\label{sec:computing_light_curves}

\begin{figure}
\centering
\includegraphics[width=\columnwidth]{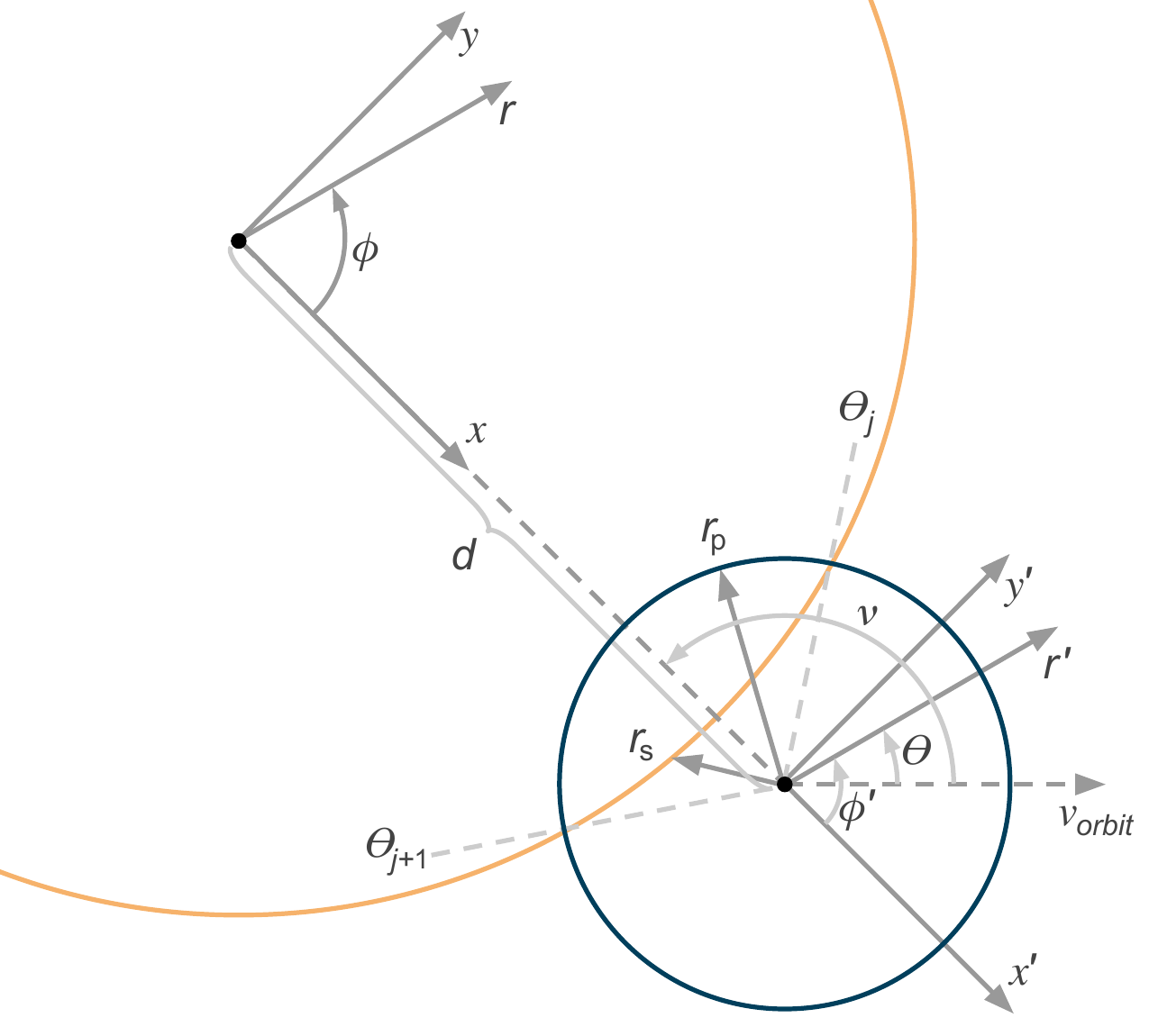}
\caption{Coordinate system definitions. Both the stellar disc (orange line) and planet's transmission string (blue line) are shown in the plane of the sky. An index of symbols can be found in Tables \ref{tab:symbols} and \ref{tab:symbols_continued}.}
\label{fig:coordinates}
\end{figure}

In this section we describe the mathematics for computing light curves of transiting exoplanets, where the occulting shape of a planet -- the transmission string -- is parameterised by a Fourier series. First we define two coordinate systems, one centred on the stellar disc defined by $x, y$ and $r,\phi$ in Cartesian and polar form, respectively, and another centred on the planet's transmission string defined by $x^{\prime}, y^{\prime}$ and $r^{\prime},\phi^{\prime}$. The two coordinate systems are separated by a distance $d$, with the $x$-directions pointing along the line of centres from the star to the planet, and the $z$-directions pointing towards the observer. Transmission strings are defined in terms of the angle $\theta$, such that they have a consistent orientation with respect to the planet's sky-projected orbital velocity vector, $\bmath{v}_{\rm{orbit}}$. We define $\nu$ as the angle between $\bmath{v}_{\rm{orbit}}$ and the line of centres, and therefore $\phi^{\prime} = \theta - \nu + \pi$. These coordinate systems are shown graphically in Figure \ref{fig:coordinates}. Note that the coordinate systems update as the planet orbits. For further reference, an index of symbols used throughout this work is provided in Tables \ref{tab:symbols} and \ref{tab:symbols_continued}.

To compute the normalised light curve flux, $F = 1 - \alpha$, we must solve the integral 
\begin{equation}
\alpha = \iint I(\mu) \,dA,
\label{eq:flux_integral}
\end{equation}
where $\alpha$ is the fractional decrease in stellar flux due to an occulting planet. This integral is computed over the sky-projected area of the planet overlapping with the stellar disc, where $I$ is the normalised stellar flux as a function of $\mu = \sqrt{1 - r^2}$ due to limb darkening. This integral is not easy to solve, especially given we should like solutions to have both fast runtimes and high precisions (see Section \ref{sec:performance_benchmarks}), all the while being applicable to a variety of stellar limb-darkening laws. For the classical circular case, a variety of solutions and approximate methods have proved widely useful \citep[e.g.,][]{mandel2002analytic, gimenez2006equations, kreidberg2015batman}. However, for our non-circular case none of these methods are optimal, owing to the complex boundary of the planet's limb. Instead, it is insights from \citet{pal2012light} that our method stands on the shoulders of most. The crucial idea is to convert the double integral over the occulted area into a line integral around the boundary enclosing this area using Green's theorem. This method has been used to great effect for computing complex light curves of mutual transits \citep{pal2012light, short2018accurate}, bodies with specific surface intensity maps \citep{luger2019starry}, and for improving the runtime and precision of the classical circular transits \citep{agol2020analytic}. In the following subsections we detail our solutions using this method, with a particular emphasis on finding solutions that are performant. For these solutions we prefer to work with transmission strings reparameterised into complex form, such that
\begin{equation}
r_{\rm{p}}(\theta) = \sum_{n=-N_c}^{N_c} c_{n} e^{i n \theta}, \\
\label{eq:transmission_string_complex}
\end{equation}
where
\begin{equation}
c_{n} = \begin{cases}
            \frac{1}{2}(a_n + i b_n), & n < 0, \\
            a_n, & n = 0, \\
            \frac{1}{2}(a_n - i b_n), & n > 0.
        \end{cases} \\
\label{eq:transmission_string_coomplex_coeffs}
\end{equation}

\subsection{Limb darkening basis}
\label{subsec:limb_darkening_basis}

We search for solutions to Equation \ref{eq:flux_integral} supporting the most common stellar limb-darkening laws: the quadratic law \citep{kopal1950detailed}, the square-root law, \citep{diaz1992new}, and the non-linear law \citep{claret2000new}. Given the overlap in terms between the square-root and non-linear law, these three parameterisations are encompassed by
\begin{align}
\label{eq:limb_darkening_law_quadratic}
& \frac{I_{\rm{q}}}{I_{\rm{q}, 0}} = 1 - u_1 (1 - \mu) - u_2 (1 - \mu)^2, \\
\label{eq:limb_darkening_law_non_linear}
& \frac{I_{\rm{nl}}}{I_{\rm{nl}, 0}} = 1 - u_1 (1 - \mu^{\frac{1}{2}}) - u_2 (1 - \mu) - u_3 (1 - \mu^{\frac{3}{2}}) - u_4 (1 - \mu^2),
\end{align}
where $u_i$ are the limb-darkening parameters, the subscripts q and nl denote the quadratic and non-linear laws, respectively, and the subscript 0 is a normalisation constant such that the total unocculted flux is unity.

Following \citet{luger2019starry} and \citet{agol2020analytic}, we find it useful to recast the limb-darkening laws as dot products of vectors, and then transform into a more convenient basis for solving the integral. This may be written
\begin{equation}
\frac{I}{I_0} = \tilde{\bmath{u}}^T \bmath{u},
\label{eq:limb_darkening_dot_product}
\end{equation}
where $\tilde{\bmath{u}}$ is the limb-darkening basis and $\bmath{u}$ is the vector of limb-darkening parameters. These vectors take the form 
\begin{equation}
\tilde{\bmath{u}}_{\rm{q}} \hspace{0.5mm} = \begin{bmatrix}
1 \\
-(1 - \mu) \\
-(1 - \mu)^2
\end{bmatrix}, 
\hspace{5.6mm}
\bmath{u}_{\rm{q}} = \begin{bmatrix}
1 \\
u_1 \\
u_2
\end{bmatrix},
\label{eq:limb_darkening_vectors_quadratic}
\end{equation}
or
\begin{equation}
\tilde{\bmath{u}}_{\rm{nl}} = \begin{bmatrix}
1 \\
-(1 - \mu^{\frac{1}{2}}) \\
-(1 - \mu) \\
-(1 - \mu^{\frac{3}{2}}) \\
-(1 - \mu^2)
\end{bmatrix},
\hspace{4.35mm}
\bmath{u}_{\rm{nl}} = \begin{bmatrix}
1 \\
u_1 \\
u_2 \\
u_3 \\
u_4
\end{bmatrix},
\label{eq:limb_darkening_vectors_non_linear}
\end{equation}
for each of the quadratic and non-linear laws, respectively. Now, Equation \ref{eq:limb_darkening_dot_product} may be written in terms of a new basis. We find a polynomial basis aids the computation of our integral, and so we have
\begin{equation}
\frac{I}{I_0} = \tilde{\bmath{p}}^T \bmath{p} = \tilde{\bmath{p}}^T \mathbfss{B} \bmath{u},
\label{eq:limb_darkening_dot_product_change_of_basis}
\end{equation}
where $\tilde{\bmath{p}}$ is the new polynomial basis, $\bmath{p}$ is the vector of limb-darkening parameters in this new basis, and $\mathbfss{B}$ is the change of basis matrix. Again, for each of the quadratic and non-linear laws, these vectors and matrices take the form
\begin{equation}
\tilde{\bmath{p}}_{\rm{q}} \hspace{0.5mm} = \begin{bmatrix}
1 \\
\mu \\
\mu^2
\end{bmatrix}, 
\hspace{5.6mm}
\mathbfss{B}_{\rm{q}} = \begin{bmatrix}
1 & -1 & -1\\
0 & 1 & 2 \\
0 & 0 & -1
\end{bmatrix},
\label{eq:limb_darkening_change_of_basis_quadratic}
\end{equation}
or
\begin{equation}
\tilde{\bmath{p}}_{\rm{nl}} = \begin{bmatrix}
1 \\
\mu^{\frac{1}{2}} \\
\mu \\
\mu^{\frac{3}{2}} \\
\mu^2
\end{bmatrix},
\hspace{4.35mm}
\mathbfss{B}_{\rm{nl}} = \begin{bmatrix}
1 & -1 & -1 & -1 & -1 \\
0 & 1 & 0 & 0 & 0 \\
0 & 0 & 1 & 0 & 0 \\
0 & 0 & 0 & 1 & 0 \\
0 & 0 & 0 & 0 & 1 \\
\end{bmatrix}.
\label{eq:limb_darkening_change_of_basis_non_linear}
\end{equation}
Using this framework, we may rewrite Equation \ref{eq:flux_integral} in terms of a vector of integrals, each element of which only depends on one power of $\mu$:
\begin{align}
\alpha &= I_0 \iint \tilde{\bmath{u}}^T \bmath{u} \,dA \nonumber \\ 
&= I_0 \iint \tilde{\bmath{p}}^T \mathbfss{B} \bmath{u} \,dA \nonumber \\
&= I_0 \bmath{s}^T \mathbfss{B} \bmath{u} \vphantom{\iint},
\label{eq:limb_darkening_integral_vector_form}
\end{align}
where
\begin{equation}
\bmath{s} = \iint \tilde{\bmath{p}} \,dA.
\label{eq:solution_integral}
\end{equation}
Our task is now to solve the integrals in Equation \ref{eq:solution_integral}. Once solutions are found for each vector element, i.e. all half-integer powers of $\mu$ arising from Equations \ref{eq:limb_darkening_change_of_basis_quadratic} and \ref{eq:limb_darkening_change_of_basis_non_linear}, we will be able to compute $\alpha$ for any of the limb-darkening laws specified above.

\subsection{The occultation integral and Green's theorem}
\label{subsec:the_occultation_integral}

Let us now consider methods of solving Equation \ref{eq:solution_integral}. As shown by \citet{pal2012light}, solutions may have favourable properties when converting these double integrals into line integrals, using Green’s theorem. We too find this approach to be beneficial, enabling us to identify closed-form solutions for some of the terms, and well-behaved approximations for others. To accomplish this conversion, we require a specific form for the $n$th element of $\tilde{\bmath{p}}$:
\begin{equation}
\tilde{p}_n = \frac{\partial D_n}{\partial x} - \frac{\partial G_n}{\partial y}.
\label{eq:pn_greens_form}
\end{equation}
Here we have defined $D_n$ and $G_n$ for which any functions satisfying $\tilde{p}_n = \mu^n$ may be chosen. \citet{short2018accurate} derive a method for finding $D_n$ and $G_n$, and tabulate these solutions (see their appendix B and table 1). Informed by the pattern of solutions for polynomial basis terms, we are able to write down the general form:

\begin{align}
D_n &= x \zeta_n, \\
G_n &= -y \zeta_n,
\label{eq:pn_greens_terms}
\end{align}
where
\begin{equation}
\zeta_n = \frac{1 - \mu^{n + 2}}{(n + 2) (1 - \mu^2)}.
\label{eq:pn_greens_zeta}
\end{equation}
 Now applying Green's theorem to Equation \ref{eq:solution_integral}, the $n$th element of $\bmath{s}$ becomes
\begin{equation}
s_n = \iint \Big( \frac{\partial D_n}{\partial x} - \frac{\partial G_n}{\partial y} \Big) \,dx dy = \oint (D_n \, dy + G_n \, dx),
\label{eq:pn_greens_integral}
\end{equation}
where the integral on the right-hand side is to be computed piecewise anti-clockwise around the closed boundary enclosing the occulted area.

\subsection{Planet-star limb intersections and the bounding curve}
\label{subsec:planet_star_limb_intersections}

To solve the line integral in Equation \ref{eq:pn_greens_integral} we must first determine its limits. This in turn requires us to determine the bounding curve around the occulted area. In general, this bounding curve is made up piecewise from sections of the stellar limb and the planet's transmission string. The exact composition of this bounding curve depends on the relative position of the planet and star, as well as the shape of the transmission string. For example, when the planet is completely overlapping with the stellar disc, the occulted area is bounded totally by the planet's transmission string. But, during ingress or egress, the occulted area is bounded by a combination of the stellar limb and the planet's transmission string. Whilst determining this bounding curve for classical circular transits has a simple solution, found from the intersection points of two circles, the solution is far less trivial when the planet's limb is parameterised by a Fourier series. In fact for complex transmission strings, there are possible configurations where the planet's limb criss-crosses the stellar limb multiple times.

To determine the composition of the bounding curve, we search for intersections between the planet's transmission string and the stellar limb in terms of the planet-centred angle, $\theta$. This amounts to finding the roots of the intersection equation
\begin{equation}
T^{\pm}(\theta) = r_{\rm{p}} - r_{\rm{s}}^{\pm} = 0,
\label{eq:intersection_equation}
\end{equation}
where $r_{\rm{s}}^{\pm}$ is the equation of an off-centred circle, i.e. the stellar limb from the planet-centred coordinate system, and the plus or minus correspond to the two potential intersection points with this circle for a given $\theta$. Substituting in the full equations and moving $r_{\rm{s}}^{\pm}$ to the right-hand side we have
\begin{equation}
\begin{split}
& \sum_{n=0}^{N_c} a_{n} \cos{(n \theta)} + \sum_{n=1}^{N_c} b_{n} \sin{(n \theta)} \\ & = d \cos{(\theta - \nu)} \pm \big(d^2 \cos^2{(\theta - \nu)} - d^2 + 1 \big)^{\frac{1}{2}}.
\end{split}
\label{eq:intersection_equation_real}
\end{equation}
To find the $\theta$s that solve this equation we convert the trigonometric functions into their complex exponential form, 
\begin{equation}
\begin{split}
\sum_{n=-N_c}^{N_c} c_{n} e^{i n \theta} = & \frac{d}{2} \big(e^{i (\theta - \nu)} + e^{-i (\theta - \nu)} \big) \\ & \pm \bigg(\frac{d^2}{4} \big(e^{i (\theta - \nu)} + e^{-i (\theta - \nu)} \big)^2 - d^2 + 1 \bigg)^{1/2},
\end{split}
\label{eq:intersection_equation_complex}
\end{equation}
and then rearrange, square, and expand, yielding
\begin{equation}
\begin{split}
& \sum_{n=-N_c}^{N_c} \sum_{m=-N_c}^{N_c} c_{n} c_{m} e^{(2N_c + n + m) i \theta} \\ & - d e^{-i \nu} \sum_{n=-N_c}^{N_c} c_{n} e^{(2N_c + n + 1) i \theta} \\ & - d e^{i \nu} \sum_{n=-N_c}^{N_c} c_{n} e^{(2N_c + n - 1) i \theta} + \big(d^2 - 1 \big)e^{2N_c i \theta} = 0.
\end{split}
\label{eq:intersection_equation_exponential_polynomial}
\end{equation}
Here we have also multiplied through by $e^{2N_c i \theta}$ to generate only positive exponents for $n > 0$. The point being that this equation is now an exponential polynomial in $e^{i \theta}$ of degree $4N_c$. Using the substitution $w = e^{i \theta}$, as suggested by \citet{weidner1988durand} for these types of problems, the equation can be written as an algebraic polynomial,
\begin{equation}
H(w) = h_0 + h_1 w + \dots + h_{4N_c - 1} w^{4N_c - 1} + h_{4N_c} w^{4N_c},
\label{eq:intersection_equation_algebraic_polynomial}
\end{equation}
where the coefficients are
\begin{equation}
h_{j} = 
    \begin{cases}
      \sum_{n = -N_c}^{-N_c + j} c_{n} c_{\rho}, \hfill \hspace{0.8mm}\scalebox{0.9}{\rm{$0 \leq j < N_c - 1$}} \hphantom{,} \\[1ex]
      \sum_{n = -N_c}^{-N_c + j} c_{n} c_{\rho} - d e^{i \nu} c_{\xi}, \hfill \hspace{0.8mm}\scalebox{0.9}{\rm{$N_c - 1 \leq j < N_c + 1$}} \hphantom{,} \\[1ex]
      \sum_{n = -N_c}^{-N_c + j} c_{n} c_{\rho} - d e^{i \nu} c_{\xi} - d e^{-i \nu} c_{\chi}, \hfill \hspace{0.8mm}\scalebox{0.9}{\rm{$N_c + 1 \leq j < 2N_c$}} \hphantom{,} \\[1ex]
      \sum_{n = -N_c}^{+N_c} c_{n} c_{\rho} - d e^{i \nu} c_{\xi} - d e^{-i \nu} c_{\chi} + d^2 - 1, \hfill \hspace{0.8mm}\scalebox{0.9}{\rm{$j = 2N_c$}} \hphantom{,} \\[1ex]
      \sum_{n = -3N_c + j}^{+N_c} c_{n} c_{\rho} - d e^{i \nu} c_{\xi} - d e^{-i \nu} c_{\chi}, \hfill \hspace{0.8mm}\scalebox{0.9}{\rm{$2N_c + 1 \leq j < 3N_c$}} \hphantom{,} \\[1ex]
      \sum_{n = -3N_c + j}^{+N_c} c_{n} c_{\rho} - d e^{-i \nu} c_{\chi}, \hfill \hspace{0.8mm}\scalebox{0.9}{\rm{$3N_c \leq j < 3N_c + 2$}} \hphantom{,} \\[1ex]
      \sum_{n = -3N_c + j}^{+N_c} c_{n} c_{\rho}, \hfill \hspace{0.8mm}\scalebox{0.9}{\rm{$3N_c + 2 \leq j < 4N_c + 1$}},
    \end{cases} \nonumber \\
\end{equation}
and the indexes $\rho = j - n - 2N_c$, $\xi = j - 2N_c + 1$, and $\chi = j - 2N_c - 1$. Polynomials, such as this, have well-trodden methods for finding their roots. In particular, \citet{boyd2006computing} shows how the roots of a Fourier series may be found by linear algebra operations. It turns out the roots of a monic polynomial are equivalent to the eigenvalues of the so-called Frobenius companion matrix, $\mathbfss{C}$. For our intersection problem, the companion matrix is
\begin{equation}
\mathbfss{C}(H) =
    \begin{bmatrix}
        0 & 0 & \dots & 0 & -h_0/h_{4N_c} \\
        1 & 0 & \dots & 0 & -h_1/h_{4N_c} \\
        0 & 1 & \dots & 0 & -h_2/h_{4N_c} \\
        \vdots & \vdots & \ddots & \vdots & \vdots \\
        0 & 0 & \dots & 1 & -h_{4N_c - 1}/h_{4N_c}
    \end{bmatrix},
\label{eq:companion_matrix}
\end{equation}
where the elements are
\begin{equation}
C_{jk} = 
    \begin{cases}
      \delta_{j, k + 1}, \hfill \hspace{12.5mm}\scalebox{0.9}{\rm{$1 \leq j < 4N_c + 1, 1 \leq k < 4N_c$}} \hphantom{,} \\[1ex]
      (-1) \frac{h_{j - 1}}{h_{4N_c}}, \hfill \hspace{12.5mm}\scalebox{0.9}{\rm{$1 \leq j < 4N_c + 1, k = 4N_c$}},
    \end{cases}
\label{eq:companion_matrix_elements}
\end{equation}
and $\delta_{j, k + 1}$ is the Kronecker delta function. So for a given configuration, we construct $\mathbfss{C}$ and compute its eigenvalues, $w_j$. These complex values solve Equation \ref{eq:intersection_equation_algebraic_polynomial} for $H(w_j) = 0$. To convert back to $\theta$ we invert the substitution,
\begin{equation}
\theta_j = -i \ln{(w_j)} = -i (ln{|w_j|} + i \arg{(w_j)} + 2 \pi n),
\label{eq:intersection_polynomial_complex_thetas}
\end{equation}
but since we only require intersections that have purely real angles in the principal branch, 
\begin{equation}
\theta_j = \arg{(w_j)},
\label{eq:intersection_polynomial_real_thetas}
\end{equation}
where $|w_j| = 1$. This result implies real intersections of the planet's transmission string and the stellar limb occur when the arguments of the eigenvalues of $\mathbfss{C}$ lie on the unit circle in the complex plane. These intersections are sorted from $-\pi$ to $\pi$ and are allocated to a vector, $\bmath{\theta}$, of length $N_{\theta}$. So that these angles span the complete closed bounding curve of the occulted area, the first element of $\bmath{\theta}$ is duplicated, $2\pi$ is added, and then appended to the end of $\bmath{\theta}$. The adjacent angles form the limits of the piecewise line integral in Equation \ref{eq:pn_greens_integral}.

For each piecewise integral, in addition to the limits, we must also determine if the bounding curve between $\theta_j$ and $\theta_{j + 1}$ is along the stellar limb or the planet's transmission string. We define each piece of the bounding curve to have a type $L(\theta_j, \theta_{j + 1}) \in \{ \mlq r_{\rm{p}} \mrq, \mlq r_{\rm{s}}^{\pm} \mrq \}$. The type is deduced by checking which intersection equation the angles solve, $T^{+}$ or $T^{-}$, and the derivatives of this intersection equation at these angles.
The logic for determining $L$ is described in full in Appendix \ref{appendix:bounding_curve_composition}, including for the trivial configurations where no intersections are found and $\bmath{\theta} = (-\pi, \pi)^T$. 

\subsection{Line integral solutions}
\label{subsec:line_integral_solutions}

Our solutions to the line integral in Equation \ref{eq:pn_greens_integral} are categorised by the type of bounding curve segment, $L(\theta_j, \theta_{j + 1})$, and the element of $\bmath{s}$ being solved. This may be written as
\begin{equation}
s_n = \sum_{j=1}^{N_{\theta}} 
    \begin{cases}
    \vphantom{s_{n \in \frac{1}{2}\mathbb{Z}, \rm{p}}} s_{n, \rm{s}}(\theta_j, \theta_{j + 1}), \hfill \hspace{2.1mm}\scalebox{0.9}{\rm{$L(\theta_j, \theta_{j + 1}) = \mlq r_{\rm{s}} \mrq, n \in \{0, \frac{1}{2}, 1, \frac{3}{2}, 2\}$}} \hphantom{,} \\[1ex]
    \vphantom{s_{n \in \frac{1}{2}\mathbb{Z}, \rm{p}}} s_{n \in 2\mathbb{Z}, \rm{p}}(\theta_j, \theta_{j + 1}), \hfill \hspace{2.1mm}\scalebox{0.9}{\rm{$L(\theta_j, \theta_{j + 1}) = \mlq r_{\rm{p}} \mrq, n \in \{0, 2\}$}} \hphantom{,} \\[1ex]
    s_{n \in \frac{1}{2}\mathbb{Z}, \rm{p}}(\theta_j, \theta_{j + 1}), \hfill \hspace{2.1mm}\scalebox{0.9}{\rm{$L(\theta_j, \theta_{j + 1}) = \mlq r_{\rm{p}} \mrq, n \in \{\frac{1}{2}, 1, \frac{3}{2}\}$}},
    \end{cases} \hspace{-5mm}
\label{eq:solution_cases}
\end{equation}
where $s_{n, \rm{s}}$ are the solutions along the stellar limb and $s_{n, \rm{p}}$ are the solutions along the planet's transmission string. The $s_{n, \rm{p}}$ solutions are further subdivided into even polynomial basis exponents, $n \in 2\mathbb{Z}$, and odd and half-integer polynomial basis exponents, $n \in \frac{1}{2}\mathbb{Z}$.

\subsubsection{Stellar limb segments}
\label{subsubsec:line_integral_solutions_stellar_limb}

For segments along the stellar limb, we know that $\mu$ is always
\begin{equation}
\mu_{\rm{s}}(\theta) = 0.
\label{eq:z_coord_stellar_limb}
\end{equation}
Substituting this into Equations \ref{eq:pn_greens_zeta} and \ref{eq:pn_greens_integral}, along with the Cartesian equations and differential elements for along the stellar limb (from Appendix \ref{appendix:coordinates_and_symbols}), we find
\begin{equation}
\begin{split}
s_{n, \rm{s}}(\theta_j, \theta_{j + 1}) & = \int_{\phi_j}^{\phi_{j + 1}} \frac{1}{n + 2} \, d\phi \\ & = \frac{1}{n + 2} (\phi_{j + 1} - \phi_j),
\end{split}
\label{eq:solution_stellar_limb}
\end{equation}
where
\begin{equation}
\phi_j = \arctan{\bigg(\frac{-r_{\rm{p}}(\theta_j) \sin{(\theta_j - \nu)}}{-r_{\rm{p}}(\theta_j) \cos{(\theta_j - \nu)} + d}\bigg)}.
\label{eq:solution_stellar_limb_phi}
\end{equation}

\subsubsection{Planet transmission string segments: $n \in 2\mathbb{Z}$}
\label{subsubsec:line_integral_solutions_planet_limb_even_terms}

For segments along the planet's transmission string, $\mu$ is given by
\begin{equation}
\mu_{\rm{p}}(\theta) = \big( 1 - d^2 - r_{\rm{p}}^2(\theta) + 2 d \cos{(\theta - \nu)} r_{\rm{p}}(\theta) \big)^{\frac{1}{2}}.
\label{eq:z_coord_planet_limb}
\end{equation}
Again, let us substitute this into Equations \ref{eq:pn_greens_zeta} and \ref{eq:pn_greens_integral}, along with the Cartesian equations and differential elements for along the planet's transmission string (from Appendix \ref{appendix:coordinates_and_symbols}). This results in the expression 
\begin{equation}
s_{n, \rm{p}}(\theta_j, \theta_{j + 1}) = \int_{\theta_j}^{\theta_{j + 1}} \zeta_n (\mu_{\rm{p}}(\theta)) \eta(\theta) \, d\theta,
\label{eq:solution_planet_limb_integral}
\end{equation}
where we have defined
\begin{equation}
\eta(\theta) = r_{\rm{p}}^2(\theta) - d \cos{(\theta - \nu)} r_{\rm{p}}(\theta) -d \sin{(\theta - \nu)} \frac{d r_{\rm{p}}(\theta)}{d\theta}.
\label{eq:solution_planet_limb_integral_eta}
\end{equation}
This integral is rather more difficult to solve than Equation \ref{eq:solution_stellar_limb}, and in fact we only find closed-form solutions for $n \in 2\mathbb{Z}$. For these terms the integral in full is
\begin{equation}
\begin{split}
& s_{n \in 2\mathbb{Z}, \rm{p}}(\theta_j, \theta_{j + 1}) = \int_{\theta_j}^{\theta_{j + 1}} \frac{1}{(n + 2)} \\ & \vphantom{} \times \Big(\sum_{k=0}^{\frac{n}{2}} \big(1 - d^2 - r_{\rm{p}}^2(\theta) + 2 d \cos{(\theta - \nu)} r_{\rm{p}}(\theta) \big)^k \Big) \\ & \times \Big( r_{\rm{p}}^2(\theta) - d \cos{(\theta - \nu)} r_{\rm{p}}(\theta) -d \sin{(\theta - \nu)} \frac{d r_{\rm{p}}(\theta)}{d\theta} \Big) \, d\theta.
\label{eq:solution_planet_limb_integral_full}
\end{split}
\end{equation}
We are able to solve this integral by realising that each term may be represented as a Fourier series in the parameter $\theta$. Terms with no $\theta$ dependence may be written as Fourier series of zeroth order, trigonometric terms in $\theta$ may be written as Fourier series up to first order, and $r_{\rm{p}}$ terms are already in this representation. These Fourier series are then combined by a succession of sums or products, wherein the series coefficient vectors are summed or convolved, respectively. The result of this process is an integrand comprised of a single new Fourier series, which can be readily solved:
\begin{equation}
\begin{split}
s_{n \in 2\mathbb{Z}, \rm{p}}(\theta_j, \theta_{j + 1}) & = \int_{\theta_j}^{\theta_{j + 1}} \sum_{m=-N_q}^{N_q} q_{n, m} e^{i m \theta} \, d\theta \\ & = \sum_{m=-N_q}^{N_q} \frac{q_{n, m}}{i m} \Big(e^{i m \theta_{j + 1}} - e^{i m \theta_j} \Big),
\end{split}
\label{eq:solution_planet_limb_integral_done}
\end{equation}
where,
\begin{equation}
\begin{split}
q_{n, m} = \frac{1}{n + 2} \Bigg[ & \Big(\sum_{k=0}^{\frac{n}{2}} \big((1 - d^2)\bmath{e}_0 - \bmath{c} * \bmath{c} + 2 d \bmath{\beta}_{\rm{cos}} * \bmath{c} \big)^{* k} \Big) \\ & * \Big( \bmath{c} * \bmath{c} - d \bmath{\beta}_{\rm{cos}} * \bmath{c} - d \bmath{\beta}_{\rm{sin}} * (\bmath{\Delta} \circ \bmath{c}) \Big) \Bigg]_m ,
\end{split}
\label{eq:solution_planet_limb_integral_coeffs}
\end{equation}
and $m$ denotes the coefficient vector element. The range of $m$ runs from $-N_q$ to $N_q$, where $N_q = \frac{n + 2}{4} (\max{(4N_c + 1, 2N_c + 3)} - 1)$. $\bmath{c}$ is a vector representation of the complex Fourier coefficients. The symbols $*$ and $* k$ represent the convolution and convolution power operators, respectively. We also define $\bmath{e}_0$, $\bmath{\Delta}$, $\bmath{\beta}_{\rm{sin}}$, and $\bmath{\beta}_{\rm{cos}}$ in Equation \ref{eq:solution_planet_limb_integral_coeffs}, all of which are detailed in Appendix \ref{appendix:coordinates_and_symbols}. Also note that for Equation \ref{eq:solution_planet_limb_integral_done}, the term in the summation takes the value $q_{n, 0} (\theta_{j + 1} - \theta_j )$ in the limit $m=0$.

\subsubsection{Planet transmission string segments: $n \in \frac{1}{2}\mathbb{Z}$}
\label{subsubsec:line_integral_solutions_planet_limb_odd_terms}

For the integral in Equation \ref{eq:solution_planet_limb_integral}, where $n \in \frac{1}{2}\mathbb{Z}$, we compute an approximate solution. We are unable to find a closed-form solution owing to half-integer exponents appearing in the $\zeta$ term. These exponents preclude the Fourier series from being combined as per the technique described in Section \ref{subsubsec:line_integral_solutions_planet_limb_even_terms}. To approximate the solution we employ Gauss–Legendre quadrature, yielding the equation
\begin{equation}
s_{n \in \frac{1}{2}\mathbb{Z}, \rm{p}}(\theta_j, \theta_{j + 1}) \approx \frac{\theta_{j + 1} - \theta_j}{2} \sum_{k=1}^{N_l} \zeta_n(\mu_{\rm{p}}(t_k)) \eta(t_k) \gamma_k,
\label{eq:solution_planet_limb_gl_approx}
\end{equation}
where
\begin{equation}
t_k = \frac{\theta_{j + 1} - \theta_j}{2} (\omega_k + 1) + \theta_j.
\label{eq:solution_planet_limb_gl_approx_mod_roots}
\end{equation}
Here $N_l$ is the number of terms to use in the approximation, $\gamma_k$ are the associated weights, and $\omega_k$ are the roots of the $N_l$-th Legendre polynomial. 

\subsection{Computation summary}
\label{subsec:computation_summary}

Putting together the derivations from the previous sections, we can now rewrite an equation for the normalised light curve flux as
\begin{equation}
F = 1 - \alpha = 1 - I_0 \bmath{s}^T \mathbfss{B} \bmath{u},
\label{eq:summary_flux}
\end{equation}
where the elements of $\bmath{s}$ are
\begin{equation}
s_n = \sum_{j=1}^{N_{\theta}} 
    \begin{cases}
    \vphantom{\sum_{m=-N_q}^{N_q}} \frac{1}{n + 2} (\phi_{j + 1} - \phi_j), \hfill \hspace{1.5mm}\scalebox{0.9}{\rm{$L = \mlq r_{\rm{s}} \mrq, n \in 2\mathbb{Z}, \frac{1}{2}\mathbb{Z}$}} \hphantom{,} \\[2ex]
    \sum_{m=-N_q}^{N_q} \frac{q_{n, m}}{i m} \Big(e^{i m \theta_{j + 1}} - e^{i m \theta_j} \Big), \hfill \hspace{1.5mm}\scalebox{0.9}{\rm{$L = \mlq r_{\rm{p}} \mrq, n \in 2\mathbb{Z}$}} \hphantom{,} \\[2ex]
    \vphantom{\sum_{m=-N_q}^{N_q}} \frac{\theta_{j + 1} - \theta_j}{2} \sum_{k=1}^{N_l} \zeta_n(\mu_{\rm{p}}(t_k)) \eta(t_k) \gamma_k, \hfill \hspace{1.5mm}\scalebox{0.9}{\rm{$L = \mlq r_{\rm{p}} \mrq, n \in \frac{1}{2}\mathbb{Z}$}},
    \end{cases} \hspace{-15mm}
\label{eq:summary_sn}
\end{equation}
and the intersections, $\bmath{\theta}$, and bounding curve types $L(\theta_j, \theta_{j + 1})$ are determined by the logic described in Section \ref{subsec:planet_star_limb_intersections}.

\section{Performance benchmarks}
\label{sec:performance_benchmarks}

We implement the mathematics for computing our transit light curves in C++, and provide an open-source python package, \texttt{harmonica}\footnote[4]{https://github.com/DavoGrant/harmonica}, for interfacing with our algorithm. The aim of this algorithm is to solve Equation \ref{eq:summary_flux}, for any user-defined transmission string and planetary orbit, at high precision and with a fast runtime. It is important that the precision of our algorithm is well beyond that of any data we wish to model, and the runtime is sufficiently fast such that this method may be incorporated, without a huge time and energy cost, into frameworks for regression modelling.

\begin{figure}
\centering
\includegraphics[width=\columnwidth]{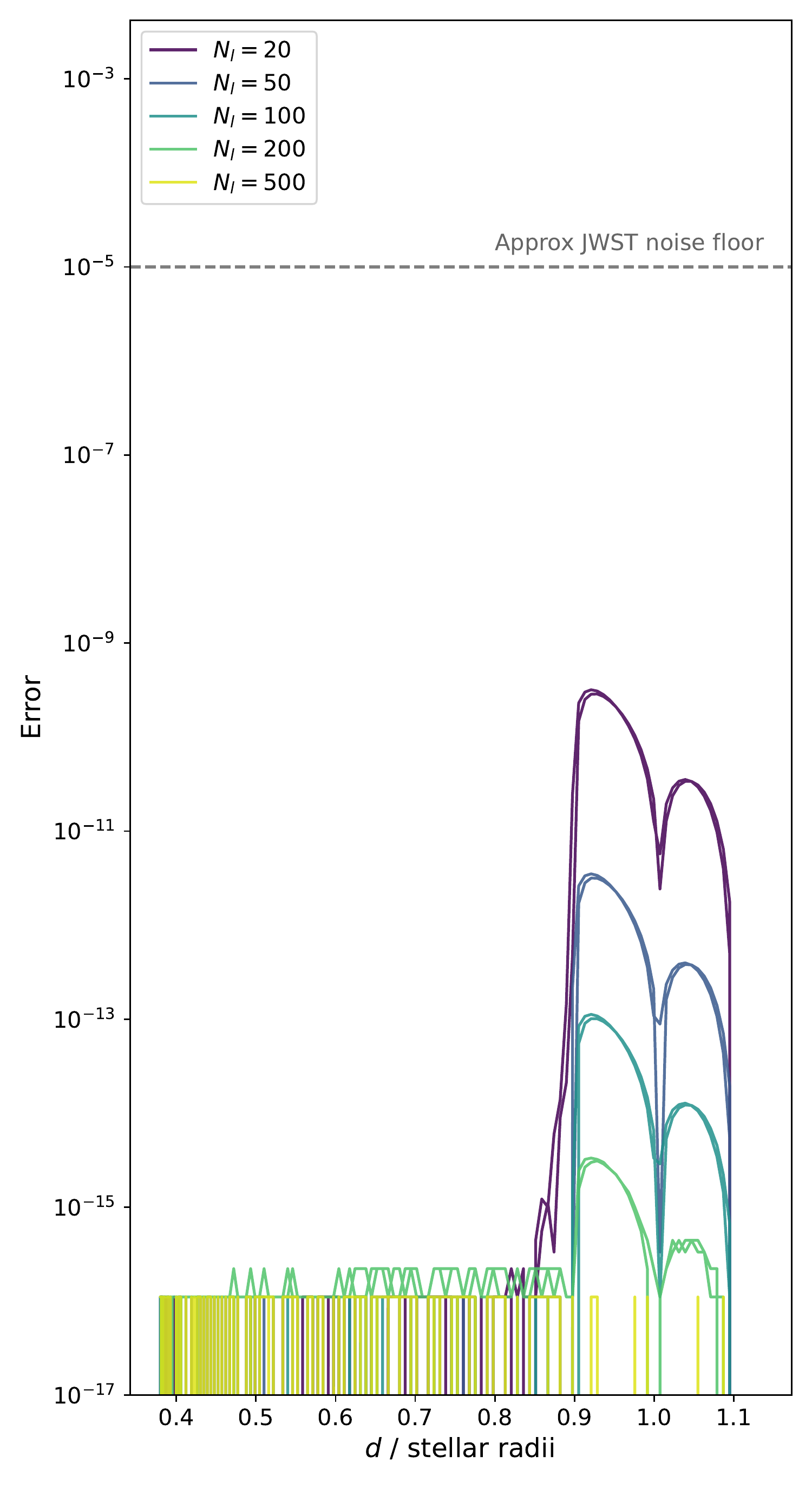}
\caption{Benchmarking the precision of a typical transit light curve, with a five-term transmission string, as a function of planet-star separation, $d$. The injected transmission string has a mean radius 0.1 stellar radii, and ${\sim}1\%$ deviations from a circular shape. The tests are conducted for various values of $N_l$.}
\label{fig:benchmark_precision}
\end{figure}

\begin{figure}
\centering
\includegraphics[width=\columnwidth]{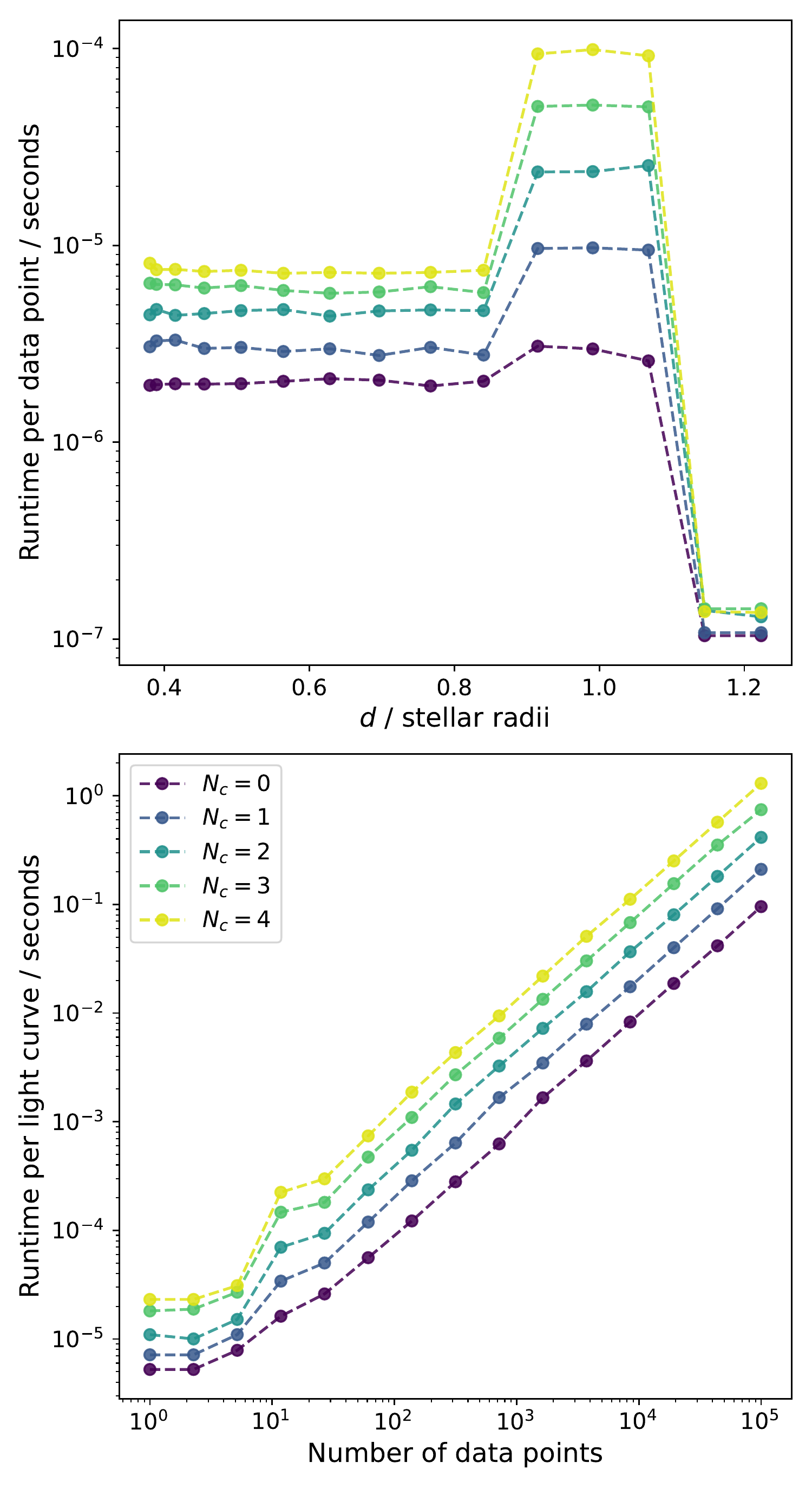}
\caption{Benchmarking the runtime of a typical transit light curve, with a five-term transmission string, as a function of planet-star separation, $d$ (top panel), and number of data points (bottom panel). The tests are conducted for various values of $N_c$, where the number of transmission string terms is equal to $2N_c + 1$. Runtimes are taken as the average over 100 tests.}
\label{fig:benchmark_runtime}
\end{figure}

\begin{figure}
\centering
\includegraphics[width=\columnwidth]{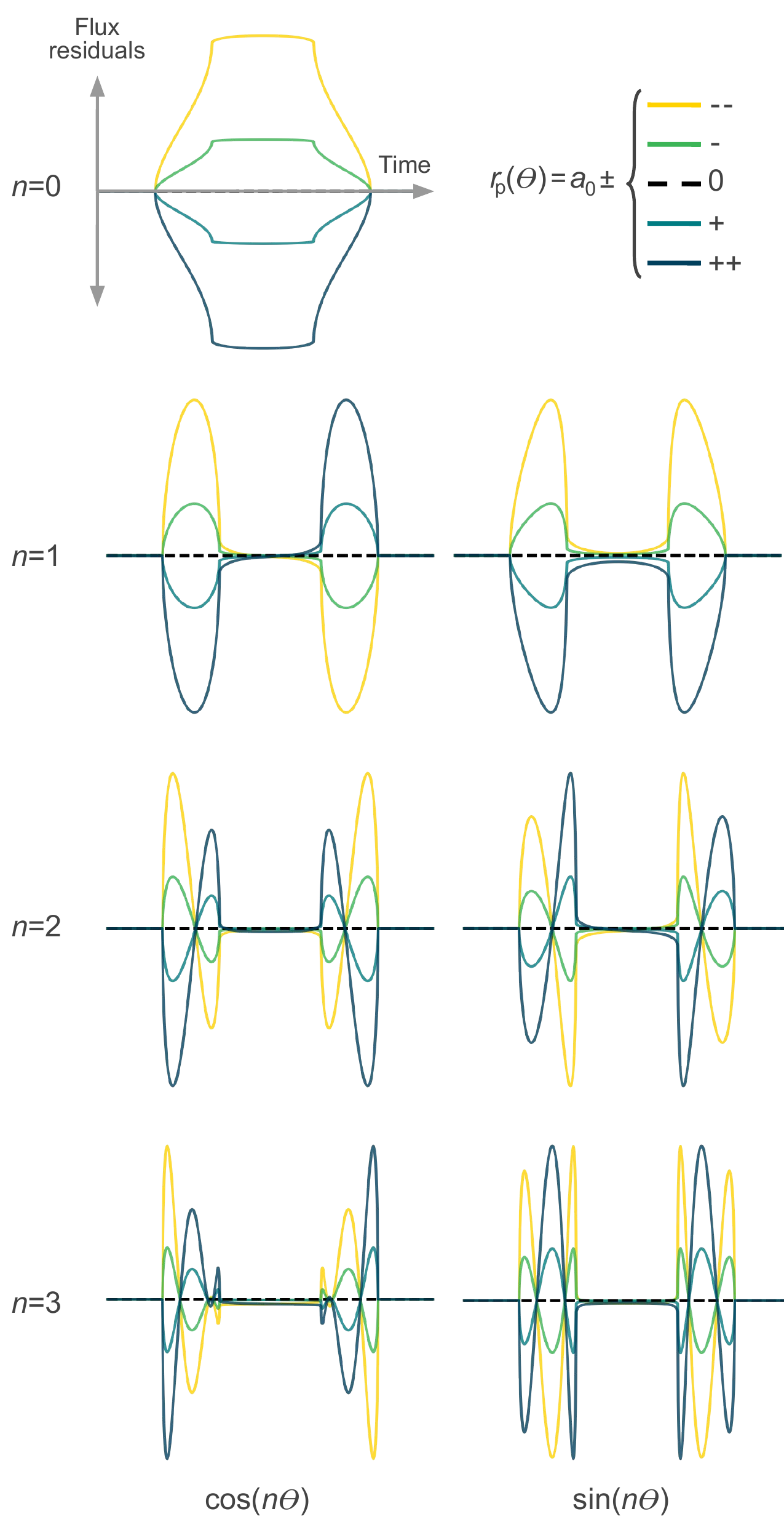}
\caption{Transit light curve residuals for the first 7 terms of our Fourier parameterisation of transmission strings. Light curves are generated for a transmission string composed of the zeroth-order term, plus or minus the corresponding grid point's high-order term at varying amplitudes. These light curves have a classical circular light curve subtracted, and the resulting residuals are shown. For reference, this figure follows the shapes in Figure \ref{fig:fourier_terms_schematic}.}
\label{fig:fourier_terms_residuals}
\end{figure}

\subsection{Precision}
\label{subsec:precision}

To assess the precision of our algorithm we compare the generated transit light curves against those from a high-precision numerical integrator, \texttt{scipy.integrate} \citep[][]{2020SciPy-NMeth}, for various transmission strings. The intrinsic precision of the numerical integrator is approximately double-precision (${\sim}10^{-16}$). For purely circular transmission strings we compare to analytic light curves, \texttt{exoplanet-core} \citep{foreman2021exoplanet}, which are also computed to double-precision \citep{agol2020analytic}. 

Our algorithm has one tunable parameter, $N_l$, which determines the number of terms used to approximate $s_{n \in \frac{1}{2}\mathbb{Z}, \rm{p}}$, and we explore suitable values for this parameter. We test a wide range of orbits, limb darkening, and transmissions strings, checking the light curve precision. We find the precision of our algorithm is orders of magnitude more precise than the data of present day observing facilities. As a demonstrative case, in Figure \ref{fig:benchmark_precision} we present the results from a typical test. We inject a transmission string with five parameters, generating ${\sim}1\%$ deviations from a circular shape with a mean radius of 0.1 stellar radii. We test values of $N_l$ ranging from 20 to 500, and find a steady progression in minimum precision value from $10^{-9}$ to $10^{-16}$. The precision shows a strong dependence on the separation of the planet and stellar centres, with configurations having the transmission string traverse the very edge of the stellar disc proving the hardest to approximate. This can be seen in Figure \ref{fig:benchmark_precision} by the double peaked features between separations of 0.9 and 1.1 stellar radii.

To balance precision verses runtime, we set default values of $N_l$ for configurations when the planet is completely inside the stellar disc, $N_l=20$, and for configurations when the planet intersects the stellar limb, $N_l=50$. Although we make these values tunable for users who wish to check their model precision. We conduct one final test of precision, where we set the quadratic limb-darkening parameters to values that satisfy the equation $u_1 + 2 u_2 = 0$. In this case the $s_{n=1, \rm{p}}$ term disappears, the resulting light curve has a purely analytic solution, and the precision reaches double-precision for all separations.

\subsection{Runtime}
\label{subsec:runtime}

The runtime\footnote[3]{All runtimes are assessed on one 2 GHz Intel core i5 processor.} of our algorithm is assessed in two ways. First, we time the computation of individual light curve data points as a function of planet-star separation. In the top panel of Figure \ref{fig:benchmark_runtime}, we present the results for a typical transiting system for a range of values of $N_c$. Note that the number of transmission string terms is equal to $2 N_c + 1$. As expected, the runtime increases as the number of transmission string terms increases. For a given value of $N_c$, the highest runtime occurs when the planet's transmission string is intersecting with the stellar limb. This jump in runtime results from having to solve the eigenvalue problem described in Section \ref{subsec:planet_star_limb_intersections}. Solving for these eigenvalues has a time-complexity that is approximately cubic with the number of transmission string terms, and as a result, this operation becomes the main bottleneck to our algorithm. Although, we are able to restrict this operation to occur only when the planet is in the vicinity of the stellar limb, by initially computing the maximum transmission string radius, and then checking at each data point if the intersection algorithm is required.

For the second test, we time the computation of entire light curves comprised of varying numbers of data points. In the bottom panel of Figure \ref{fig:benchmark_runtime}, we present the results for a typical transiting system, again, for a range of values of $N_c$. The runtime shows an approximately linear dependence on the number of data points. We observe a slight bump to longer runtimes for light curves of about 10 data points. These bumps are due to the light curves having sufficient data points that some of these points intersect with the stellar limb. Overall, for a typical light curve of 1000 data points, and a transmission string comprised of five parameters, the runtime is approximately 4ms. 

\subsection{Derivatives}
\label{subsec:derivatives}

Model optimisation and inference may be aided by computation of derivatives with respect to input parameters. For example, in least-squares regression via the Jacobian \citep[e.g.,][]{more1978levenberg}, or in Bayesian inference using Hamiltonian Monte Carlo (HMC) \citep[][]{duane1987hybrid, neal2011mcmc}. To this end, we derive model derivatives with respect to each input parameter. The derived equations are supplied in the online supplementary material.

\section{Demonstration}
\label{sec:demo}

\begin{figure*}
\centering
\includegraphics[width=0.94\textwidth]{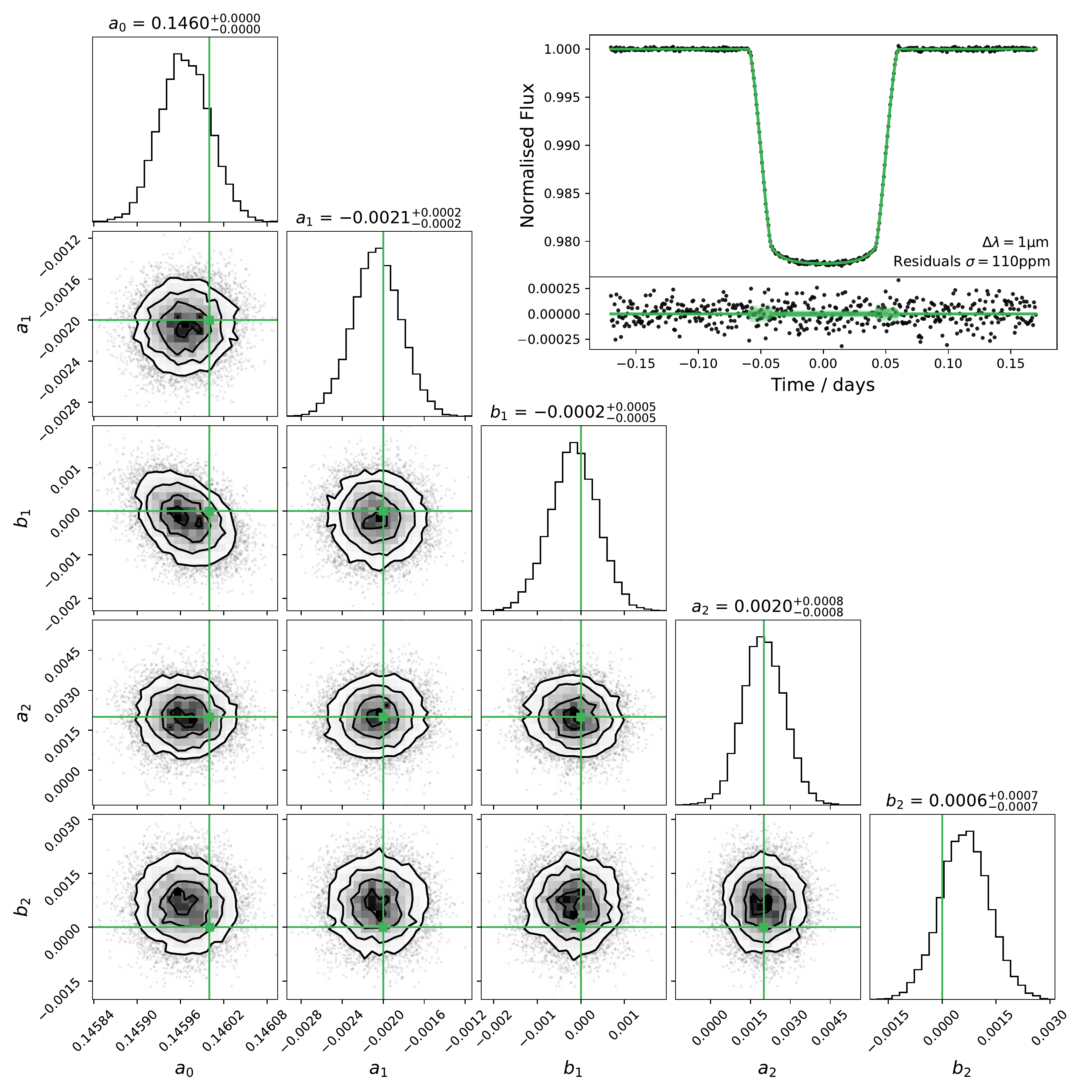}
\caption{Bottom-left panels: posterior distributions and covariances of a 5-parameter transmission string fitted to a JWST-like transit light curve. The injected parameters values are shown by the green lines. Top-right: the upper panel shows light curves drawn from the posterior distributions plotted against the simulated data. The lower panel shows the residuals between these light curve draws and the simulated data.}
\label{fig:demo_corner}
\end{figure*}

\begin{figure*}
\centering
\includegraphics[width=\textwidth]{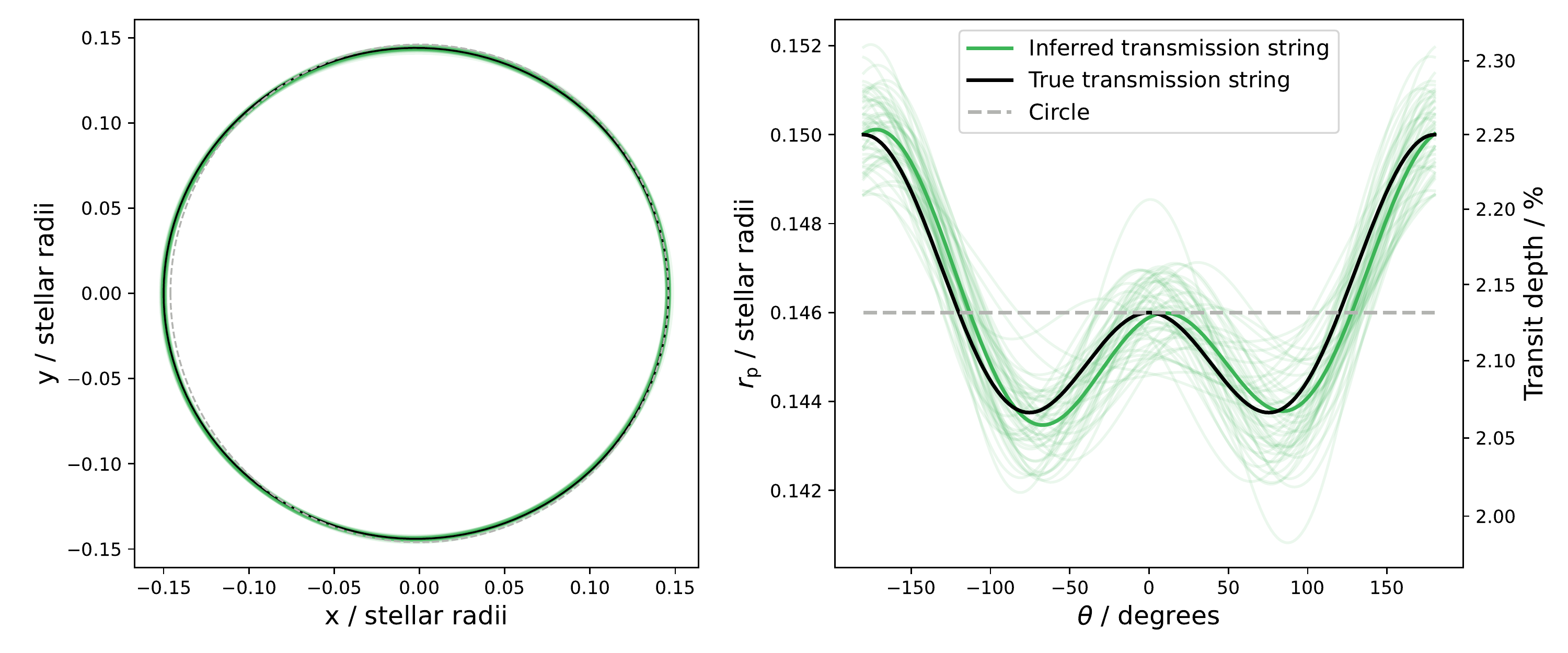}
\caption{Inferences of a 5-parameter transmission string fit to a JWST-like transit light curve. The median transmission string (green), transmission string samples from the posterior distributions (light green), and true injected transmission string (black) are shown relative to a circle (grey dashes). The left panel shows the sky-projected shape, while the right panel shows the transmission string as a function of angle around the terminator.}
\label{fig:demo_inferred}
\end{figure*}

To demonstrate the capabilities of our algorithm, in this section we generate a set of forward models, highlighting the signals imprinted in the transit data by various transmission string terms. We then simulate a JWST-like dataset for a typical hot Jupiter, and present the inferences we are able to make about the shape of the transmission string relative to the injected shape.

\subsection{Forward models}
\label{subsec:forward_models}

We generate transit light curves for a set of planet transmission strings. Following the shapes presented in Figure \ref{fig:fourier_terms_schematic}, we start with a circular transmission string and then add or subtract small amounts of the higher-order Fourier terms one at a time. We compare each of the generated transit light curves to the purely circular case, and display the residuals in Figure \ref{fig:fourier_terms_residuals}. All the light curves are generated for identical planetary orbits, which have their inclinations set to $87^{\rm{\circ}}$. 

We find each Fourier term imprints a unique signal into the light curve residuals. The harmonic order determines how many distinct peaks there are in the residuals, and the harmonic type, either cosine or sine, determines the order of these peaks at ingress and egress. For example, the $\cos{(2 \theta)}$ term shows two residual peaks at both ingress and egress, with the order of the peaks mirrored about the transit centre. In contrast, the $\sin{(2 \theta)}$ term shows the same two peaks, but in this case the residual peaks are mirrored and flipped about the transit centre. Overall, the forward models encode distinguishable shapes that enable the inverse procedure: inferring the shapes directly from the light curves.

\subsection{Inference}
\label{subsec:inference}

A JWST-like transit light curve is simulated for a typical hot Jupiter. The orbit and mean planet radius are based on WASP-39b \citep{fischer2016hst, mancini2018gaps}, and we inject a known 5-parameter transmission string with parameters $a_0=0.146$, $a_1=-0.002$, $b_1=0.0$, $a_2=0.002$, and $b_2=0.0$. This transmission string corresponds to an atmosphere with an inflated east-versus-west terminator, as well as an inflated equatorial-versus-polar region.

We elect to simulate data for this system using JWST's NIRSpec/G395H mode, which spans a wavelength range from 2.7 to 5.2 \textmu m and provides high information content specifically for exoplanet atmosphere characterisation \citep{batalha2017information}. To simulate a realistic cadence and noise level we use \texttt{pandexo} \citep{batalha2017pandexo}. The resulting light curves are comprised of 465 integrations over the course of a 0.34 day observation. From these data we select a 1 \textmu m interval, centred on 3.2 \textmu m, to generate a single high-quality light curve. This is representative of a binning scheme aimed at probing the spatial distribution of molecular features in the infrared. Lastly, the star is given quadratic limb darkening calculated using \texttt{exotic-ld} \citep[][]{Laginja2020, hannah_wakeford_2022_6809899}. The resulting simulated light curve is shown in the top right of Figure \ref{fig:demo_corner}. This light curve has white noise with a standard deviation of 110 ppm.

Next, we demonstrate our algorithm's inference capabilities by fitting the simulated transit light curve for a 5-parameter transmission string. The orbital and limb-darkening parameters are held fixed at their known simulated values. The fitting is performed using an HMC Markov Chain Monte Carlo sampler from the \texttt{NumPyro} package \citep{phan2019composable, bingham2019pyro}. We run the sampler in two parallel chains, each for 5000 steps, and discard the first 2000 steps as warm-up.

In Figure \ref{fig:demo_corner} we show the model fit (top-right) and the marginalised posteriors and covariances for the 5 transmission string parameters (bottom-left panels). All 5 of the injected transmission string parameters are recovered to within 1 sigma of their simulated values. We also find that the posterior distributions appear Gaussian, and any covariance between the transmission string parameters is minimal. In Figure \ref{fig:demo_inferred} we show realisations of the inferred transmission string from the parameter posterior distributions. The transmission string is constrained to a standard deviation 0.56\% of the mean (circular) radius, easily enabling the injected shape to be resolved.

This demonstration shows the fidelity with which transmission strings may be extracted directly from light curve data. Of course the constraints we may place on the inferred transmission strings are related to the quality of the data. As such, our recommended methodology is to start by fitting only the zeroth-order harmonic, corresponding to the classical circular case, and gradually add higher-order harmonics to the transmission string. The number of harmonics may then be justified through a model selection process; for example, by comparing models by their marginalised likelihoods or by some approximate information criterion. In this way, one may allow the data to inform the complexity of the transmission strings.

In this demonstration we have inferred one transmission string from one light curve. By repeating this process for multiple wavelength channels, a set of transmission strings may then be used to generate transmission spectra as a function of angle around the planet's terminator (recall Figure \ref{fig:transmission_string_schematic}).

\section{Discussion}
\label{sec:discussion}

The technique developed in this work opens up the ability to model transit light curves with adjustable complexity. The Fourier parameterisation allows for terms, or harmonics, to be added as the data requires. Starting with just one term, the zeroth-order harmonic, and our model is equivalent to that of the classical circular case. Next, by including two more terms, the first harmonic, and our model reaches similar complexity to that of the two-hemispheres model of \citet{EspinozaJones2021AJ}. Further harmonics may then be included to generate more and more complex shapes. The complexity of our parameterisation is only limited by the constraint that the radius must be a single-valued function of angle around the terminator. For an even more general framework of transiting shapes, including variable opacity levels, see the technique developed by \citet{sandford2019shadow}.

For high-precision observations, our technique enables spatial information around a planet's terminator to be drawn out at the light curve fitting stage. This may be particularly important for subsequent retrieval analysis, as it has been shown that multidimensional effects can lead to degeneracies between clouds and metallicity \citep{LineParmentier2016ApJ}, as well as temperature biases \citep{MacDonald2020}. Our technique may be able to extract some of these multidimensional signals, and then allow retrievals to be performed for a transmission spectrum at a given angle around the terminator.

An additional process worth noting is that of a planet's rotation. Any rotation between ingress and egress may lead to variation in the sky-projected shape of the occultor. As shown by \citet{wardenier2022all}, for many planets this rotation is smaller than the absorption region probed by transit data. However, for some ultra-hot Jupiters this may not be the case. For these close-in planets, the absorption region may change significantly between ingress and egress, and as such we may also expect changes to the shape of the transmission strings. To model these cases we include additional flexibility in the provided code, allowing for time-dependent transmission string parameterisations. In order to not immensely increase the number of free parameters, it may be judicious to define a transmission string at first contact, a transmission string at fourth contact, and interpolate between the values throughout transit.

The intended use case for our technique is as a tool for studying exoplanet atmospheres. However, the technique can be applied more generally to any transit light curve, where the shape of the transiting body is of interest.

\section{Summary and conclusions}
\label{sec:summary_and_conclusions}
In this study we have devised a technique for spatially mapping exoplanet atmospheres in transmission. The primary utility of such a technique is in probing the multidimensional physics at play in exoplanet atmospheres. Our work into this new technique is summarised as follows:
\begin{enumerate}

\item We defined a mathematical object, referred to as a
{\it transmission string}, that describes the planet radius as a single-valued function of the angle around the planet's terminator. These objects enable spatial information encoded in transit light curves to be directly extracted, with one light curve yielding one transmission string. Given observations of light curves at multiple wavelengths, all the inferred transmission strings may be used to generate a transmission spectrum at any angle around the planet's terminator.

\item The transmission strings were parameterised in terms of Fourier series. This choice was motivated by Fourier series generating practical shapes with only a small number of terms. The first few harmonics naturally produce shapes that can probe differences in the east-to-west, north-to-south, and equatorial-vs-polar radii. Furthermore, this parameterisation has the flexibility to build transmission strings of arbitrary complexity, all the while being reducible to the classical circular case.

\item We devised the mathematics to compute transit light curves of planetary transmission strings. The emphasis of our formulation was on performance. The final algorithm is precise, fast, and differentiable, and so our model may be easily incorporated into frameworks for regression.

\item We demonstrated the inference capabilities of our technique with simulated data. We synthesised a JWST-like transit light curve and ran an injection and recovery test of a known transmission string. Our results showed how high-fidelity spatial information may be inferred, given that high-precision light curves are available, such as those from JWST.

\end{enumerate}
The technique presented in this work is provided as an open-source package, called \texttt{harmonica}. The source code is available at https://github.com/DavoGrant/harmonica, with documentation and tutorials hosted at https://harmonica.readthedocs.io.

\section*{Acknowledgements}
\label{sec:acknowledgements}
We would like to thank the referee for a helpful and constructive report. We also thank N. E. Batalha, T. J. Wilson, L. Alderson, M. Lodge, R. J. MacDonald, D. Rindt, and R. R. Surgenor for helpful discussions. We gratefully acknowledge the use of the following software: numpy \citep[][]{harris2020array}, SciPy \citep[][]{2020SciPy-NMeth}, matplotlib \citep[][]{Hunter:2007}, exotic-ld \citep[][]{Laginja2020, hannah_wakeford_2022_6809899}, corner \citep[][]{foreman2016corner}, pybind11 \citep[][]{jakob2017pybind11}, eigen v3 \citep[][]{guennebaud2010eigen}, jax \citep[][]{jax2018github}, and numpyro \citep[][]{phan2019composable, bingham2019pyro}. D. Grant acknowledges funding from the UKRI STFC Consolidated Grant ST/V000454/1. 

\section*{Data availability}
\label{sec:data_availability}
There are no new data associated with this article.




\bibliographystyle{mnras}
\bibliography{harmonica_refs} 



\appendix

\section{Coordinates and symbols}
\label{appendix:coordinates_and_symbols}

The coordinate systems used throughout this work are depicted in Figure \ref{fig:coordinates}. There are two coordinate systems, one centred on the stellar disc and another centred on the planet’s transmission string, separated by a distance $d$. 

The Cartesian equations and differential elements for coordinates along the stellar limb are
\begin{align}
    \label{eq:cartesian_x_stellar_limb}
    & x_{\rm{s}} = \cos{\phi}, \\   
    \label{eq:cartesian_dx_stellar_limb}
    & dx_{\rm{s}} = -\sin{\phi} d\phi, \\   
    \label{eq:cartesian_y_stellar_limb}
    & y_{\rm{s}} = \sin{\phi}, \\
    \label{eq:cartesian_dy_stellar_limb}
    & dy_{\rm{s}} = \cos{\phi} d\phi,
\end{align}
and for along the planet's transmission string are
\begin{align}
    \label{eq:cartesian_x_planet_limb}
    & x_{\rm{p}} = r_{\rm{p}}(\theta) \cos{\phi^{\prime}} + d = -r_{\rm{p}}(\theta) \cos{(\theta - \nu)} + d \vphantom{\frac{d r_{\rm{p}}(\theta)}{d\theta}}, \\   
    \label{eq:cartesian_dx_planet_limb}
    & dx_{\rm{p}} = -\cos{(\theta - \nu)} \frac{d r_{\rm{p}}(\theta)}{d\theta} d\theta + r_{\rm{p}}(\theta) \sin{(\theta - \nu)} d\theta , \\   
    \label{eq:cartesian_y_planet_limb}
    & y_{\rm{p}} = r_{\rm{p}}(\theta) \sin{\phi^{\prime}} = -r_{\rm{p}}(\theta) \sin{(\theta - \nu)} \vphantom{\frac{d r_{\rm{p}}(\theta)}{d\theta}}, \\
    \label{eq:cartesian_dy_planet_limb}
    & dy_{\rm{p}} = -\sin{(\theta - \nu)} \frac{d r_{\rm{p}}(\theta)}{d\theta} d\theta - r_{\rm{p}}(\theta) \cos{(\theta - \nu)} d\theta .
\end{align}

In Equation \ref{eq:solution_planet_limb_integral_coeffs} we define several vectors for brevity of notation. $\bmath{e}_n$ are vectors of unit length in the $n$th elements direction. For example, $\bmath{e}_0$ is a vector that spans indexes from $-N_c$ to $N_c$. This vector is full of zeroes, except the $0$th position is a one. Additionally, we also define the vectors
\begin{equation}
\bmath{\Delta} = \sum_{n=-N_c}^{N_c} i n \bmath{e}_n, 
\label{eq:solution_planet_limb_integral_coeffs_delta}
\end{equation}
\begin{equation}
\bmath{\beta}_{\rm{sin}} = \cos{\nu} \begin{bmatrix}
          \frac{i}{2}  \\
          \hspace{1.4mm}0\hspace{1.4mm} \\
          -\frac{i}{2}
         \end{bmatrix}
      - \sin{\nu} \begin{bmatrix}
          \frac{1}{2}  \\
          \hspace{1.4mm}0\hspace{1.4mm} \\
          \frac{1}{2}
         \end{bmatrix},
\label{eq:solution_planet_limb_integral_coeffs_beta_sin}
\end{equation}
\begin{equation}
\bmath{\beta}_{\rm{cos}} = \cos{\nu} \begin{bmatrix}
          \frac{1}{2}  \\
          \hspace{1.4mm}0\hspace{1.4mm} \\
          \frac{1}{2}
         \end{bmatrix}
      + \sin{\nu} \begin{bmatrix}
          \frac{i}{2}  \\
          \hspace{1.4mm}0\hspace{1.4mm} \\
          -\frac{i}{2}
         \end{bmatrix},
\label{eq:solution_planet_limb_integral_coeffs_beta_cos}
\end{equation}
which are each used for combing the Fourier series with various coefficients.

In Tables \ref{tab:symbols} and \ref{tab:symbols_continued} we provide an index of symbols.

\begin{table}
    \centering
    \begin{tabular}{l p{50mm} l}
    \hline
    Symbol & Definition & Reference \\
    \hline
    $a_n$ & Cosine $n$th harmonic amplitude & Equation
    \ref{eq:transmission_string_real} \\
    $b_n$ & Sine $n$th harmonic amplitude & Equation
    \ref{eq:transmission_string_real} \\
    $\mathbfss{B}$ & Generic change of basis matrix & Equation
    \ref{eq:limb_darkening_dot_product_change_of_basis} \\
    $\mathbfss{B}_{\rm{q}}$ & Quadratic change of basis matrix & Equation
    \ref{eq:limb_darkening_change_of_basis_quadratic} \\
    $\mathbfss{B}_{\rm{nl}}$ & Non-linear change of basis matrix & Equation
    \ref{eq:limb_darkening_change_of_basis_non_linear} \\
    $c_n$ & Complex $n$th harmonic amplitude & Equation
    \ref{eq:transmission_string_coomplex_coeffs} \\
    $\bmath{c}$ & Complex harmonic amplitudes vector & Equation
    \ref{eq:solution_planet_limb_integral_coeffs} \\
    $C_{jk}$ & Frobenius companion matrix elements & Equation
    \ref{eq:companion_matrix_elements} \\
    $\mathbfss{C}$ & Frobenius companion matrix & Equation
    \ref{eq:companion_matrix} \\
    $d$ & Separation of stellar and planet centres & Section
    \ref{sec:computing_light_curves} \\
    $D_n$ & $n$th element of the anti-derivative wrt $x$ in \newline\indent\hspace{3.5mm} the polynomial basis & Equation
    \ref{eq:pn_greens_form} \\
    $\bmath{e}_n$ & Unit vector in $n$th direction & Appendix
    \ref{appendix:coordinates_and_symbols} \\
    $F$ & Normalised transit light curve flux & Section
    \ref{sec:computing_light_curves} \\
    $G_n$ & $n$th element of the anti-derivative wrt $y$ in \newline\indent\hspace{3.5mm} the polynomial basis & Equation
    \ref{eq:pn_greens_form} \\
    $h_j$ & Intersection polynomial coefficients & Equation
    \ref{eq:intersection_equation_algebraic_polynomial} \\
    $H$ & Intersection polynomial equation & Equation
    \ref{eq:intersection_equation_algebraic_polynomial} \\
    $i$ & Unit imaginary number, $\sqrt{-1}$ &  \\
    $I$ & Normalised stellar flux & Equation
    \ref{eq:flux_integral} \\
    $I_{\rm{q}}$ & Normalised quadratic limb darkening law & Equation
    \ref{eq:limb_darkening_law_quadratic} \\
    $I_{\rm{nl}}$ & Normalised non-linear limb darkening law & Equation
    \ref{eq:limb_darkening_law_non_linear} \\
    $I_0$ & Normalised constant $I(\mu=1)$ & Equation
    \ref{eq:limb_darkening_dot_product} \\
    $I_{\rm{q, 0}}$ & Normalisation constant, $I_{\rm{q}}(\mu=1)$ & Equation
    \ref{eq:limb_darkening_law_quadratic} \\
    $I_{\rm{nl, 0}}$ & Normalisation constant, $I_{\rm{nl}}(\mu=1)$ & Equation
    \ref{eq:limb_darkening_law_non_linear} \\
    $L$ & Bounding curve type & Appendix
    \ref{appendix:bounding_curve_composition} \\   
    $N_c$ & Number of transmission string terms is equal \newline\indent\hspace{3.5mm} to $2N_c + 1$ & Equation
    \ref{eq:transmission_string_real} \\
    $N_{\theta}$ & Number of planet-star intersections & Section
    \ref{subsec:planet_star_limb_intersections} \\
    $N_q$ & Number of combined Fourier series terms & Equation
    \ref{eq:solution_planet_limb_integral_done} \\
    $N_l$ & Number of Gauss–Legendre roots & Equation
    \ref{eq:solution_planet_limb_gl_approx} \\
    $\tilde{p}_n$ & $n$th element in polynomial basis & Equation
    \ref{eq:pn_greens_form} \\
    $\tilde{\bmath{p}}$ & Generic polynomial basis & Equation
    \ref{eq:limb_darkening_dot_product_change_of_basis} \\
    $\tilde{\bmath{p}}_{\rm{q}}$ & Quadratic polynomial basis & Equation
    \ref{eq:limb_darkening_change_of_basis_quadratic} \\
    $\tilde{\bmath{p}}_{\rm{nl}}$ & Non-linear polynomial basis & Equation
    \ref{eq:limb_darkening_change_of_basis_non_linear} \\
    $\bmath{p}$ & Generic polynomial parameters & Equation
    \ref{eq:limb_darkening_dot_product_change_of_basis} \\
    $\bmath{p}_{\rm{q}}$ & Quadratic polynomial parameters & Equation
    \ref{eq:limb_darkening_change_of_basis_quadratic} \\
    $\bmath{p}_{\rm{nl}}$ & Non-linear polynomial parameters & Equation
    \ref{eq:limb_darkening_change_of_basis_non_linear} \\
    $q_{n, m}$ & Combined Fourier series coefficients & Equation
    \ref{eq:solution_planet_limb_integral_coeffs} \\
    $r$ & Polar coordinate in stellar-centred frame & Section
    \ref{sec:computing_light_curves} \\
    $r^{\prime}$ & Polar coordinate in planet-centred frame & Section
    \ref{sec:computing_light_curves} \\
    $r_{\rm{p}}$ & Transmission string/planet limb & Equation
    \ref{eq:transmission_string_real} \\
    $r_{\rm{s}}$ & Stellar limb & Equation
    \ref{eq:intersection_equation_real} \\
    $s_n$ & $n$th element in solution vector & Equation
    \ref{eq:pn_greens_integral} \\
    $s_{n, \rm{s}}$ & Line segment solution along $r_{\rm{s}}$ & Equation
    \ref{eq:solution_cases} \\
    $s_{n \in 2\mathbb{Z}, \rm{p}}$ & Line segment solution along $r_{\rm{p}}$ for even \newline\indent\hspace{3.5mm} polynomial basis exponents & Equation
    \ref{eq:solution_cases} \\
    $s_{n \in \frac{1}{2}\mathbb{Z}, \rm{p}}$ & Line segment solution along $r_{\rm{p}}$ for odd and \newline\indent\hspace{3.5mm} half-integer polynomial basis exponents & Equation
    \ref{eq:solution_cases} \\
    $\bmath{s}$ & Solution vector of integrals & Equation
    \ref{eq:solution_integral} \\
    $t_k$ & Rescaled Gauss–Legendre roots & Equation
    \ref{eq:solution_planet_limb_gl_approx_mod_roots} \\
    $T$ & Intersection equation & Equation
    \ref{eq:intersection_equation} \\
    $u_i$ & $i$th limb-darkening parameter & Equation
    \ref{eq:limb_darkening_law_quadratic} \\
    $\tilde{\bmath{u}}$ & Generic limb-darkening basis & Equation
    \ref{eq:limb_darkening_dot_product} \\
    $\tilde{\bmath{u}}_{\rm{q}}$ & Quadratic limb-darkening basis & Equation
    \ref{eq:limb_darkening_vectors_quadratic} \\
    $\tilde{\bmath{u}}_{\rm{nl}}$ & Non-linear limb-darkening basis & Equation
    \ref{eq:limb_darkening_vectors_non_linear} \\
    $\bmath{u}$ & Generic limb-darkening parameters & Equation
    \ref{eq:limb_darkening_dot_product} \\
    $\bmath{u}_{\rm{q}}$ & Quadratic limb-darkening parameters & Equation
    \ref{eq:limb_darkening_vectors_quadratic} \\
    $\bmath{u}_{\rm{nl}}$ & Non-linear limb-darkening parameters & Equation
    \ref{eq:limb_darkening_vectors_non_linear} \\
    $\bmath{v}_{\rm{orbit}}$ & Planet sky-projected orbital velocity vector & Appendix
    \ref{appendix:coordinates_and_symbols} \\
    $w$ & Intersection equation substitution variable & Equation
    \ref{eq:intersection_equation_algebraic_polynomial} \\
    $w_j$ & Eigenvalue of the companion matrix & Equation
    \ref{eq:intersection_equation_algebraic_polynomial} \\
    $x$ & Cartesian coordinate in stellar-centred frame & Figure
    \ref{fig:coordinates} \\
    \hline
    \end{tabular}
    \caption{Symbols index.}
    \label{tab:symbols}
\end{table}

\begin{table}
    \centering
    \begin{tabular}{l p{50mm}l l}
    \hline
    Symbol & Definition & Reference \\
    \hline
    $x^{\prime}$ & Cartesian coordinate in planet-centred frame & Figure
    \ref{fig:coordinates} \\
    $x_{\rm{s}}$ & Cartesian $x$-coordinate along $r_{\rm{s}}$ & Equation
    \ref{eq:cartesian_x_stellar_limb} \\
    $x_{\rm{p}}$ & Cartesian $x$-coordinate along $r_{\rm{p}}$ & Equation
    \ref{eq:cartesian_x_planet_limb} \\
    $dx_{\rm{s}}$ & Differential element $dx$ along $r_{\rm{s}}$ & Equation
    \ref{eq:cartesian_dx_stellar_limb} \\
    $dx_{\rm{p}}$ & Differential element $dx$ along $r_{\rm{p}}$ & Equation
    \ref{eq:cartesian_dx_planet_limb} \\
    $y$ &  Cartesian coordinate in stellar-centred frame & Figure
    \ref{fig:coordinates} \\
    $y^{\prime}$ & Cartesian coordinate in planet-centred frame & Figure
    \ref{fig:coordinates} \\
    $y_{\rm{s}}$ & Cartesian $y$-coordinate along $r_{\rm{s}}$ & Equation
    \ref{eq:cartesian_y_stellar_limb} \\
    $y_{\rm{p}}$ & Cartesian $y$-coordinate along $r_{\rm{p}}$ & Equation
    \ref{eq:cartesian_y_planet_limb} \\
    $dy_{\rm{s}}$ & Differential element $dy$ along $r_{\rm{s}}$ & Equation
    \ref{eq:cartesian_dy_stellar_limb} \\
    $dy_{\rm{p}}$ & Differential element $dy$ along $r_{\rm{p}}$ & Equation
    \ref{eq:cartesian_dy_planet_limb} \\
    $2\mathbb{Z}$ & The set of even integers & Section
    \ref{subsec:line_integral_solutions} \\
    $\frac{1}{2}\mathbb{Z}$ & The set of odd integers and half-integers & Section
    \ref{subsec:line_integral_solutions} \\
    $\alpha$ & Fractional occulted stellar flux & Equation
    \ref{eq:flux_integral} \\
    $\bmath{\beta}_{\rm{sin}}$ & Vector of sine modifiers & Equation
    \ref{eq:solution_planet_limb_integral_coeffs_beta_sin} \\
    $\bmath{\beta}_{\rm{cos}}$ & Vector of cosine modifiers & Equation
    \ref{eq:solution_planet_limb_integral_coeffs_beta_cos} \\
    $\gamma_k$ & Gauss–Legendre weights & Equation
    \ref{eq:solution_planet_limb_gl_approx} \\  
    $\delta$ & Kronecker delta function &  \\
    $\bmath{\Delta}$ & Vector of derivative modifiers & Equation
    \ref{eq:solution_planet_limb_integral_coeffs_delta} \\
    $\zeta_n$ & Function used in Green's theorem conversion & Equation
    \ref{eq:pn_greens_zeta} \\
    $\eta$ & Function used in solution integral & Equation
    \ref{eq:solution_planet_limb_integral_eta} \\
    $\theta$ & Transmission string coordinate & Equation
    \ref{eq:transmission_string_real} \\
    $\theta_j$ & Planet-star intersection angle, start of piece- \newline\indent\hspace{3.5mm} wise line segment in planet-centred frame & Equation
    \ref{eq:intersection_polynomial_real_thetas} \\
    $\theta_{j + 1}$ & Planet-star intersection angle, end of piece- \newline\indent\hspace{3.5mm} wise line segment in planet-centred frame & Equation
    \ref{eq:intersection_polynomial_real_thetas} \\
    $\bmath{\theta}$ & Planet-star intersections vector & Equation
    \ref{eq:transmission_string_real} \\
    $\mu$ & Limb-darkening radial parameter in stellar \newline\indent\hspace{3.5mm} -centred frame & Figure
    \ref{fig:coordinates} \\
    $\mu^{\prime}$ & Limb-darkening radial parameter in planet- \newline\indent\hspace{3.5mm} -centred frame & Figure
    \ref{fig:coordinates} \\
    $\mu_{\rm{s}}$ & Limb-darkening radial parameter along $r_{\rm{s}}$ & Equation
    \ref{eq:z_coord_stellar_limb} \\
    $\mu_{\rm{p}}$ & Limb-darkening radial parameter along $r_{\rm{p}}$ & Equation
    \ref{eq:z_coord_stellar_limb} \\
    $\nu$ & Angle between $\bmath{v}_{\rm{orbit}}$ and the line of centres & Section
    \ref{sec:computing_light_curves} \\
    $\xi$ & Intersection polynomial coefficient index & Equation
    \ref{eq:intersection_equation_algebraic_polynomial} \\
    $\rho$ & Intersection polynomial coefficient index & Equation
    \ref{eq:intersection_equation_algebraic_polynomial} \\
    $\phi$ & Polar coordinate in stellar-centred frame & Section
    \ref{sec:computing_light_curves} \\
    $\phi^{\prime}$ & Polar coordinate in planet-centred frame & Section
    \ref{sec:computing_light_curves} \\
    $\phi_j$ & Planet-star intersection angle, start of piece- \newline\indent\hspace{3.5mm} wise line segment in stellar-centred frame & Equation
    \ref{eq:solution_stellar_limb_phi} \\
    $\phi_{j + 1}$ & Planet-star intersection angle, end of piece- \newline\indent\hspace{3.5mm} wise line segment in stellar-centred frame & Equation
    \ref{eq:solution_stellar_limb_phi} \\
    $\chi$ & Intersection polynomial coefficient index & Equation
    \ref{eq:intersection_equation_algebraic_polynomial} \\
    $\omega_k$ & Gauss–Legendre roots & Equation
    \ref{eq:solution_planet_limb_gl_approx_mod_roots} \\
    \hline
    \end{tabular}
    \caption{Symbols index continued.}
    \label{tab:symbols_continued}
\end{table}

\section{Occulted area bounding curve composition}
\label{appendix:bounding_curve_composition}

The occulted area is the overlap between the planet and the stellar disc. This area is bounded piecewise by a curve composed of sections of the planet's transmission string and the stellar limb. To solve the line integral in Equation \ref{eq:pn_greens_integral}, the curve type and limits for each piecewise integral must be deduced. For each pair of adjacent intersection angles, $\theta_j$ and $\theta_{j + 1}$ (see Section \ref{subsec:planet_star_limb_intersections}), the bounding curve types can be reasoned by checking which intersection equation the angles solve, $T^{+}$ or $T^{-}$, and the derivatives of the corresponding intersection equations, given by
\begin{equation}
\begin{split}
\frac{d T^{\pm}(\theta_j)}{d \theta} = & \sum_{n=-N_c}^{N_c} i n c_{n} e^{n i \theta_j} + d \sin{(\theta_j - \nu)} \\ & \pm \frac{d^2 \sin{(\theta_j - \nu)} \cos{(\theta_j - \nu)}}{\big(d^2 \cos^2{(\theta_j - \nu)} - d^2 + 1 \big)^{1/2}}.
\end{split}
\end{equation}

In Table \ref{tab:intersection_categories} we tabulate the logic for determining the type of bounding curve, $L(\theta_j, \theta_{j + 1})$, for adjacent elements of the vector $\bmath{\theta}$. Each pair of adjacent angles are associated with either the $T^{+}$ or $T^{-}$ equation. The derivatives of this equation are then checked at these angles to be either positive or negative. Given these four bits of information, represented as a single row in Table \ref{tab:intersection_categories}, we can determine $L$ uniquely.

If the derivative at an intersection point is zero, then we decipher that this intersection root has a multiplicity of 2, and can be skipped. If no intersections are found, then the configurations may be trivially deduced by
\begin{equation}
L(\theta_0, \theta_1) = 
    \begin{cases}
      \mlq r_{\rm{p}} \mrq, \hfill &\scalebox{1.0}{\rm{$d \leq 1, r_{\rm{p}}(\theta = \nu) < 1 + d$}} \\[1ex]
      \mlq r_{\rm{s}} \mrq, \hfill &\scalebox{1.0}{\rm{$d \leq 1, r_{\rm{p}}(\theta = \nu) > 1 + d$}} \\[1ex]
     \mlq  \rm{None} \mrq, \hfill &\scalebox{1.0}{\rm{$d > 1, r_{\rm{p}}(\theta = \nu) < 1 + d$}} \\[1ex]
     \mlq  r_{\rm{s}} \mrq, \hfill &\scalebox{1.0}{\rm{$d > 1, r_{\rm{p}}(\theta = \nu) > 1 + d$}},
    \end{cases}
\label{eq:trivial_configurations}
\end{equation}
where $\bmath{\theta} = (\theta_0, \theta_1)^T = (-\pi, \pi)^T$, and None indicates there is no occulted area.

\begin{table}
    \centering
    \begin{tabular}{c c c c c}
    \hline
    $T^{\pm}(\theta_j)$ & $T^{\pm}(\theta_{j + 1})$ & $\frac{d T^{\pm}(\theta_j)}{d \theta}$ & $\frac{d T^{\pm}(\theta_{j + 1})}{d \theta}$ & $L(\theta_j, \theta_{j + 1})$ \\
    \hline
    $+$ & $+$ & $-$ & $+$ & $\mlq r_{\rm{p}} \mrq$ \\
    $+$ & $+$ & $+$ & $-$ & $\mlq r_{\rm{s}} \mrq$ \\
    
    $-$ & $-$ & $-$ & $+$ & $\mlq r_{\rm{s}} \mrq$ \\
    $-$ & $-$ & $+$ & $-$ & $\mlq r_{\rm{p}} \mrq$ \\
    
    $+$ & $-$ & $-$ & $-$ & $\mlq r_{\rm{p}} \mrq$ \\
    $+$ & $-$ & $+$ & $+$ & $\mlq r_{\rm{s}} \mrq$ \\
    $-$ & $+$ & $-$ & $-$ & $\mlq r_{\rm{s}} \mrq$ \\
    $-$ & $+$ & $+$ & $+$ & $\mlq r_{\rm{p}} \mrq$ \\
    \hline
    \end{tabular}
    \caption{Logic table for determining the bounding curve type, $L$, as a function of the adjacent angles $\theta_j$ and $\theta_{j + 1}$. The columns $T^{\pm}$ test whether the angles solve $T^{+}$, denoted by a `$+$', or $T^{-}$, denoted by a `$-$'. The columns $\frac{d T^{\pm}}{d \theta}$ test whether the derivative of the intersection equation is positive, denoted by a `$+$', or negative, denoted by a `$-$'.}
    \label{tab:intersection_categories}
\end{table}


\bsp	
\label{lastpage}
\end{document}